\documentclass{aa}
\usepackage{txfonts}
\usepackage{amssymb}
\usepackage{graphicx}
\usepackage{hyperref}
\def\etal{{et al. }}

\def\eqalign#1{\null\,\vcenter{\openup1\jot 
   \ialign{\strut\hfil$\displaystyle{##}$&$\displaystyle{{}##}$\hfil 
   \crcr#1\crcr}}\,} 

\def\degs{\ifmmode ^{\circ}\else$^{\circ}$\fi}
\def\amin{\ifmmode ^{\prime}\else$^{\prime}$\fi}
\def\asec{\ifmmode ^{\prime\prime}\else$^{\prime\prime}$\fi}
\def\fm{\hbox{$.\!\!^{\rm m}$}}            
\def\fss{\hbox{$.\!\!^{\rm s}$}}        
\def\fdg{\hbox{$.\!\!^\circ$}}          
\def\farcm{\hbox{$.\mkern-4mu^\prime$}}
\def\farcs{\hbox{$.\!\!^{\prime\prime}$}}  

\def\psg{PSR J0633+1746}
\def\gem{Geminga}
\def\psh{PSR B0656+14}

\def\widerul{\vrule height 2.5ex width 0ex depth 0ex}

\begin{document}

\title{Subaru optical observations of the two middle-aged pulsars   
PSR B0656+14 and Geminga\thanks{Based on data collected at Subaru Telescope,
which is operated by the National Astronomical Observatory of Japan.}}
\author{
Yuri A.~Shibanov\inst{1},
Sergei V.~Zharikov\inst{2},
Viktoria N.~Komarova\inst{3,4},
Nobuyuki Kawai\inst{5},
Yuji Urata\inst{6},
Alexey B.~Koptsevich\inst{1,7},
Vladimir V.~Sokolov\inst{3,4},
Shinpei Shibata\inst{8},
Noriaki Shibazaki\inst{9}}
\offprints{Yu.~Shibanov, shib@astro.ioffe.ru}
\authorrunning{Yu.~Shibanov et al.}
\titlerunning{Subaru optical observations  of PSR B0656+14 and Geminga  } 
\institute{
Ioffe Physical Technical Institute, Politekhnicheskaya 26,
St. Petersburg 194021, Russia 
\and
Observatorio Astronomico Nacional SPM, Instituto de Astronomia, 
UNAM,  Ensenada, BC, Mexico
\and
Special Astrophysical Observatory of RAS,
Karachai-Cherkessia, Nizhnij Arkhyz 369167, Russia
\and
 Isaac Newton Institute of Chile, SAO Branch, Russia
\and
Department of Physics, Tokyo Institute of Technology, 
 2-12-1 Ookayama, Muguro-ku, Tokyo 152-8551, Japan
\and
RIKEN (Institute of Physical and Chemical Research) Hirosawa, Wako, 
Saitama 351-0198, Japan
\and
 University of Toronto,
 60 St George St,
 Toronto, ON M5S1A7, Canada 
\and
Yamagata University,  Yamagata 990-8560, Japan
\and
Rikkyo University, Tokyo 171-8501, Japan
}

\date{Received  09 09 2005, accepted  19 10 2005} 

\abstract{}
{We carried out  
a deep subarcsecond 
$BRI$ imaging of the two middle-aged pulsars   
to establish their properties 
in the optical range 
}
{Astrometry and photometry  methods  
are applied to identify the
pulsars and to measure their fluxes.   
 We also  reanalyze  archival ESO/NTT and HST broadband data  
and find  that  some   published   fluxes for \gem\ were estimated inaccurately.  
The resulting  dereddened  broadband spectra 
in the near-IR-UV range are analyzed and compared 
with available  data  from the radio through  gamma-rays.} 
{Both pulsars are detected at $\ga$10$\sigma$ level.
\gem\ is for the first time reliably detected 
in the $I$ band with a magnitude of 25\fm10$\pm$0.14.  
The dereddened  spectra  of both pulsars are remarkably similar to each other  
and    show  significant flux increases towards  the far-UV and  
near-IR, and  a wide flux excess in $V$-$I$ bands. This suggests a multicomponent structure
of the optical emission.   The nonthermal power law component of the pulsar magnetospheric origin 
dominates in the most part of the  optical range. For  \psh\  it is compatible with a low energy extension   of the power law tail 
seen in  hard X-rays.  For \gem\  the respective extension  overshoots  by a factor of 100 
the nonthermal optical flux, which has   a less steep spectral slope than  in X-rays.   This implies      
a spectral break  at a photon energy   $\sim$1~keV. The  flux increases 
 towards the far-UV  are  compatible with  contributions of   the Rayleigh-Jeans parts of the blackbody
components  from whole  surfaces of the neutron stars   dominating in soft X-rays.  
The $V$-$I$ excess, which is most significant for \psh,  suggests a third 
spectral component of  still   
unidentified origin. 
 Faint, a few arcseconds in size nebulae extended perpendicular to the proper motion directions of  the pulsars, 
are seen around both objects in our deepest $I$ band images. They can be optical counterparts 
of the bow-shock head of Geminga and of the tentative pulsar wind  nebula of \psh\  observed in X-rays.}
{}                         

\maketitle
\section{Introduction}
The two nearby middle-aged  pulsars, PSR~B0656+14 
and \gem\ (\psg),  belong to a small group  
of  rotation powered isolated  neutron stars  
(NSs) whose emission is detected in a wide spectral range  spanning 
from radio  through  gamma-rays. 
Their parameters,   such as the period  ${\rm P}$ 
and its derivative ${\rm \dot P}$, characteristic age $\tau$, spin-down luminosity ${\rm \dot E}$,  
magnetic field  ${\rm B}$, are very  similar (Table~\ref{t:par}). 
This is reflected in similarities of  observational manifestations of both pulsars.  

Multiwavelength spectra of PSR~B0656+14 and  Ge\-minga  consist of  two main  components.   
The first one is a nonthermal component which is believed  
to be generated in magnetospheres of the NSs.  It  dominates  
the pulsar emission in almost the whole  observed  range  and     
its  spectrum is described  by  a power law, albeit      
with different spectral indices in different  
spectral domains. Multiwavelength  properties of this component  
are important for the study of not yet clearly 
understood  emission mechanisms  
in magnetospheres of NSs. 
The second  component dominates  in soft X-ray  
and EUV ranges where it is seen  as a strong excess  
over the nonthermal background.    
It is  well described by a blackbody spectrum and this excess  
is  thought to be associated with  thermal emission 
form cooling surfaces of the NSs  (Becker \& Tr\"{u}mper~\cite{Becker};
  Zavlin \& Pavlov~\cite{zp04a}; De Luca et al.~\cite{deluca}). 
The study  of this component  is of a crucial importance to evaluate the NS surface temperature,  
to compare it with NS cooling theories and to obtain information 
on still poorly known properties and equation 
of state of super-dense  matter in  interiors
of NSs  (e.g., Yakovlev \& Pethick \cite{yp04}). 
 Inclusion of an additional high temperature thermal component  
assumed to be originated  from  hot pulsar polar caps appears to improve  
significantly  spectral fits in X-rays  (Koptsevich et al.~\cite{K2001}; Pavlov et al.~\cite{Pav02};    
 Zavlin \& Pavlov~\cite{zp04a}; De Luca et al.~\cite{deluca}).      
  
There are,  however,  noticeable  differences  between the
two objects.  \gem\ is known as one of 
the brightest gamma-ray sources on the sky (Bignami \& Caraveo \cite{big96}), 
while \psh\ is  fainter  and only marginally detected in
gamma-rays (Ramanamurthy et al.~\cite{r96}). 
In the radio range \psh\ is a stable source with properties to be typical 
for  an ordinary radio pulsar. 
In contrast,  \gem\  has been considered  for a long time 
as a ``radio-quiet'' NS, until it was detected at rather 
low (only $\la  102.5$~MHz) frequencies 
(Kuz'min \& Losovskii \cite{kuz97}; Malofeev \& Malov  
\cite{MM97}; Shitov \& Pugachev \cite{Shi98}).  \gem\ 
has  unusually  steep radio spectrum   (Burderi et
al.~1999), and  unlike most of radio
pulsars its pulse profile is very unstable 
(Kassim \& Lazio \cite{Kassim}). 
\begin{table}[t]
\caption{Parameters of \gem\ and \psh\ (from Taylor et al. \cite{Taylor},
unless specified otherwise).}
\label{t:par}
\begin{tabular}{lll}
\hline\hline
\multicolumn{1}{r}{}&\multicolumn{1}{c}{Geminga} &\multicolumn{1}{c}{B0656+14} \widerul\\
\hline
\multicolumn{3}{l}{Observed} \widerul\\
\cline{1-1}
$\alpha$ (J2000)$^a$\dotfill         &$6^h33^m54\fss1530(28)^b$   &$6^h59^m48\fss1472(7)^c$  \widerul        \\
$\delta$ (J2000)$^a$\dotfill &$17^o46^{\prime}12\farcs909(40)^b$  &$14^o14^{\prime}21\farcs160(10)^c$\\
$\mu_\alpha^a$ (mas yr$^{-1}$)\dotfill&$138(4)^d$              &$45.47(65)^c$                 \\
$\mu_\delta^a$ (mas yr$^{-1}$)\dotfill&$97(4)^d$               &$-2.40(29)^c$                   \\
$\pi^a$ (mas)\dotfill         &$6.36(1.74)^d$          &$3.47(36)^c$                  \\
$l^e$        \dotfill                &195\fdg134              &201\fdg108            \\
$b^e$        \dotfill                &4\fdg265                &8\fdg258                      \\
$P$ (ms)     \dotfill    &237                     &385                           \\
$\dot P$ ($10^{-15}$)\dotfill  &$10.97$                 &$55.01$                       \\
$D\!M^f$ (cm$^{-3}$ pc)\dotfill &$2.9^g$                 &$14.02$               \widerul\\
\hline
\multicolumn{3}{l}{Derived} \widerul\\
\cline{1-1}
$d^{a}$  (pc) \dotfill       &$157(^{+59}_{-34})^d$   &$288(^{+33}_{-27})^c$ \widerul\\
$\tau^h$ (kyr)\dotfill        &342                     &111                           \\
$\log B$  (G)  \dotfill      &$12.21$                  &$12.67$                       \\
$\log \dot E$ (erg s$^{-1}$) \dotfill &$34.57$                 &$34.58$               \widerul\\
\hline
\\
\end{tabular}
\begin{tabular}{l}
$^a$~coordinates, proper motion, parallax, and parallax-based  \\
\hspace{2ex}distance; hereafter the numerals in  parentheses are 1$\sigma$ \\
\hspace{2ex}uncertainties referring  to last significant digits quoted;\\
$^b$~Caraveo et al.~\cite{Car98} (epoch 1995.21);\\
$^c$~Brisken et al.~\cite{brisk03};\\
$^d$~Caraveo et al.~\cite{Car96};\\
$^e$~galactic coordinates;\\
$^f$~dispersion measure;  \\
$^g$~ Malofeev \& Malov~\cite{MM97};  \\
$^h$~characteristic age $P/2\dot{P}$.       \\
\end{tabular}
\end{table}

\psh\ has a twice lower transverse velocity, ${\rm v_\perp \approx 60}$~km~s$^{-1}$
(Mignani et al.~2000; Brisken et al. \cite{brisk03}), than \gem, ${\rm v_\perp \approx 122}$~km~s$^{-1}$ (Caraveo et al. \cite{Car96}), 
and both pulsars  move in different environment, as indicated by 
their dispersion measure values (cf.~Table 1). \psh\ sits near
the center of the Monogem ring (Nousek et
al.~\cite{nousek81}), 
that is a bright $\sim 10^5$~yr expanding supernova remnant (SNR)  
likely produced at the same supernova explosion as the pulsar (Thorsett et al. \cite{thor03}).
\gem\  has not been associated with any known SNR, although based on its proper
motion and age a birth place was proposed to be located inside the Cas-Tau OB 
or the Ori OB1a stellar associations  
(Pellizza et al.~\cite{pell05}).
Recent XMM-Newton  observations of this pulsar in X-rays revealed a faint bow-shock 
structure whose tails extend up to 2\arcmin\ from the pulsar and are elongated   
along direction of its proper motion (Caraveo et al. \cite{Car03}). 
The structure has not yet been identified in the optical range. 
No reliable signs of a bow-shock nebula have been detected around more distant 
and slowly moving \psh. Some hints of a compact, 3\arcsec-4\arcsec, faint pulsar 
wind nebula (PWN) surrounding the pulsar were reported based on the X-ray images 
obtained with the Chandra/LETG/HRC-S (Marshall \& Schulz \cite{Mar02}) and with the Chandra/ACIS-S3  
(Pavlov et al. \cite{Pav02}), but they  have not  yet been
confirmed by  deeper exposures  
and have not been identified in other spectral domains. 

Both pulsars are firmly identified in the optical range by an accurate
positional coincidence of the optical counterparts with the pulsars, by
comparison of the proper motion in radio and optical ranges (Caraveo et~al.~\cite{Car96};
Mignani et al.~\cite{Mig2000}; Brisken et al.~\cite{brisk03}), and by detection of  pulsations with
the pulsar period 
in the optical (\psh: Shearer et al.~\cite{Shear97}; Kern et
al.~\cite{Kern2003}. \gem: Shearer et al.~\cite{Shear98}) and in
UV     (\psh: 
Shibanov et al.~\cite{Sh2005}. 
\gem: Kargaltsev et al.~\cite{Karg05}).

Due to faintness of the pulsar counterparts the spectral information in
the optical is restricted mainly by the broadband photometry. Published
spectral observations of \gem\  (Martin et al.~\cite{Mar98}) are too noisy to compete
with the broadband data. The broadband far-UV (FUV) 
spectra of both pulsars appear to be dominated by
the Rayleigh-Jeans  (RJ)  part of the
blackbody spectrum  extrapolated from soft X-rays
 (Mignani et al.~\cite{mcb98}; Koptsevich et al.~\cite{K2001}, hereafter 
K\cite{K2001}).  For \gem\ this is 
supported also by  FUV spectral
data recently reported by Kargaltsev et
al.~(\cite{Karg05}).   At longer wavelengths there is a significant
flux excess over the RJ extrapolation in the 
spectra of both pulsars. 
\begin{table}[t]
\caption{Log of observations of \gem\ and \psh\  
with the Subaru.} 
\begin{tabular}{lllcll}
\hline\hline
   PSR       &       UT   &Exp.   &Band   &Seeing$^a$& sec~z   \\
	     & 21 Jan 2001&sec    &       &arcsec&         \\
\hline
  Geminga    & 05:17:26.9 & 332.3 & $I$ & 0.66 & 1.56   \\
  Geminga    & 05:30:12.6 & 300.0 & $I$ & 0.66 & 1.47   \\
  Geminga    & 05:42:29.2 & 300.0 & $I$ & 0.70 & 1.39   \\
  Geminga    & 05:51:53.6 & 300.0 & $I$ & 0.65 & 1.35   \\
  Geminga    & 06:01:30.5 & 300.0 & $I$ & 0.68 & 1.30   \\
  Geminga    & 06:31:27.7 & 300.0 & $I$ & 0.55 & 1.19   \\
  Geminga    & 06:40:32.8 & 300.0 & $I$ & 0.55 & 1.16   \\
  Geminga    & 06:49:38.0 & 400.0 & $I$ & 0.55 & 1.13   \\
  Geminga    & 07:00:22.8 & 400.0 & $I$ & 0.60 & 1.11   \\
  Geminga    & 07:11:07.8 & 400.0 & $I$ & 0.58 & 1.09   \\
  Geminga    & 07:21:52.8 & 400.0 & $I$ & 0.60 & 1.07   \\
  Geminga    & 07:32:38.2 & 400.0 & $I$ & 0.60 & 1.05   \\
  Geminga    & 07:43:23.9 & 400.0 & $I$ & 0.60 & 1.04   \\
  Geminga    & 08:07:55.3 & 400.0 & $I$ & 0.61 & 1.01   \\
  Geminga    & 08:26:36.9 & 400.0 & $I$ & 0.62 & 1.00   \\
  B0656+14   & 08:46:10.7 & 400.0 & $I$ & 0.65 & 1.01   \\
  B0656+14   & 08:56:55.8 & 400.0 & $I$ & 0.65 & 1.01   \\
  B0656+14   & 09:07:40.9 & 400.0 & $I$ & 0.73 & 1.00   \\
  B0656+14   & 09:18:26.0 & 400.0 & $I$ & 0.67 & 1.00   \\
  B0656+14   & 09:29:10.9 & 400.0 & $I$ & 0.65 & 1.01   \\
  B0656+14   & 09:39:55.9 & 400.0 & $I$ & 0.63 & 1.01   \\
  B0656+14   & 10:22:05.8 & 510.0 & $R$ & 0.60 & 1.04   \\
  B0656+14   & 10:34:41.0 & 510.0 & $R$ & 0.60 & 1.06   \\
  Geminga    & 10:47:15.8 & 510.0 & $R$ & 0.61 & 1.13   \\
  Geminga    & 10:59:58.6 & 510.0 & $R$ & 0.68 & 1.17   \\
  Geminga    & 11:12:33.8 & 510.0 & $R$ & 0.61 & 1.21   \\
  Geminga    & 11:25:09.0 & 510.0 & $R$ & 0.60 & 1.25   \\
  B0656+14   & 11:54:36.9 & 600.0 & $B$     & 0.65 & 1.28   \\
  B0656+14   & 12:08:42.3 & 600.0 & $B$     & 0.66 & 1.35   \\
  B0656+14   & 12:22:47.8 & 600.0 & $B$     & 0.69 & 1.43   \\
\hline
\end{tabular}
\label{t:log}
\begin{tabular}{ll}
$^a$~FWHM of the stellar profile in the individual images.   & \\
\end{tabular}
\end{table}

For each of the pulsars these excesses have been 
interpreted in a different way.  For \psh\ the spectral energy distribution 
from the optical to near-IR (NIR) ranges  was shown to be compatible with the low 
energy extension of a high energy nonthermal spectral tail detected in X-rays 
(K\cite{K2001}). It suggested possible common
origin of the nonthermal spectral component  in the optical
and X-rays. 
Early attempts of detection of \gem\ in NIR were unsuccessful
and the measured upper limit in the $I$ band 
suggested a strong  spectral dip in this band   followed by a prominent excess in the 
$V$ band
(Mignani et al.~\cite{mcb98}).  The shape of the excess
 reminded a broad  emission feature  over the RJ
continuum that has been interpreted  
as an ion cyclotron emission  from   
a hot non-uniformly magnetized plasma covering the pulsar polar caps
 (Jacchia et al.~\cite{Jacchia}).  However, later Geminga was  detected in
the NIR  (Harlow et al.~\cite{Harlow}). Preliminary data analysis  (K\cite{K2001})
revealed a flux growth  towards longer  wavelengths, as it is in case  of \psh.
This does not quite agree  with the cyclotron interpretation. In addition,
Kurt et al.~(\cite{Kurt2001}) reported on a tentative detection of \gem\ in
the $I$ band at a flux level, which is  by a half order of magnitude higher
than the upper limit known previously. The reported flux value is compatible
with a smooth connection  between the $V$ and  NIR bands,
that has made the cyclotron interpretation even more questionable.
 
In any case, published  broadband optical data are still too uncertain to
reject or confirm surely the suggested interpretations and to understand
whether these two middle-aged pulsars have similar or different properties
in the optical range. The uncertainties do not allow us to study also 
significance of the spectral difference 
between the middle-aged
and other pulsars of different ages detected in optical
range   (K\cite{K2001}; 
Zharikov et al.~\cite{Z2002}, \cite{Z2003}; Shibanov et al.~\cite{Sh2003}). The main source of
the flux errors in case of \psh\ is a contamination of the pulsar flux by
a background extended object in $\simeq $~1\farcs1 from the pulsar 
(K\cite{K2001}). For about twice fainter \gem, there are no contaminating
objects but deeper exposures are necessary to establish 
its  optical spectral energy distribution 
with a higher accuracy.
    
In this paper we report on a subarcsecond deep broadband imaging of the 
\gem\  and \psh\
fields with the Subaru telescope. Preliminary
results have been reported by Komarova \etal (\cite{K2003}).  Here
we analyze our data together with the
available NIR-optical-UV data from the  HST, ESO/NTT, 
and 6m BTA telescope, as well as
with available multiwavelength data from the radio through 
gamma-rays.  In Sect.~\ref{s:obs} we 
describe the observations, data reduction, astrometric and photometric
referencing. The results are presented in Sect.~\ref{s:res}
and discussed in Sect.~\ref{s:dis}.
\section{Observations and data analysis}
\label{s:obs}
 \subsection{Subaru observations and data reduction}
Two fields  containing  \gem\ and PSR B0656+14 were observed on
January 21 2001 with the wide field camera Suprime-Cam at the primary focus of
the Subaru telescope. The Suprime-Cam (Miyazaki et al.~\cite{Miyazaki}) was equipped with
ten MIT/LL $2048 \times 4096$ CCDs arranged in a $5\times 2$ pattern to provide
$34^{\prime} \times 27^{\prime}$ FOV with the pixel size on the sky of 
$0.201^{\prime\prime}  \times 0.201^{\prime\prime}$.
The pulsars were exposed in one of the CCD chips, {\tt si006s}, with 
the Gain$=$1.17, making use of the $B$, $R$, and $I$ filters
with  throughput to be close to  
the Johnson-Cousins system. Sets of 5-10~min individual dithered exposures were
obtained in each of the filters to get rid of  cosmic rays 
at further  data analysis. Several  individual exposures
were removed from the consideration due to problems with the telescope guiding system. 
Thus the total  exposure time  used for the subsequent analysis of \gem\
was 2040~s  in the $R$ and 5332.3~s in the $I$ bands. For a brighter \psh\  it was  
 1800~s, 1020~s, and 2400~s in the $B$, $R$, and $I$ bands, respectively.  The observational conditions were rather
stable during the night with seeing varying from $\sim0\farcs6$ to $\sim0\farcs7$. 
 A {\sl Log} of observations is given in 
Table \ref{t:log}. The PG1047+003 standard field of Landolt (\cite{lan92}) used 
for the photometric calibration was observed at the end of the night. 
\begin{table*}[t] 
\caption{Positions of \gem\ and \psh\ at the  epoch 
of the Subaru observations, 2001.054.}
\begin{tabular}{clcllcllr}
\hline
\hline 
\multicolumn{3}{c}{Object}&\multicolumn{3}{c}{Expected}&\multicolumn{2}{c}{Observed}& \\  
&&&\multicolumn{1}{c}{$\alpha_{2000}$}&\multicolumn{1}{c}{$\delta_{2000}$}&&
\multicolumn{1}{c}{$\alpha_{2000}$}&\multicolumn{1}{c}{$\delta_{2000}$}&\\
\hline
&Geminga&& 06$^h$33$^m$54$\fss$2067(32) &17$^o$46$'$13\farcs475(46)&
&06$^h$33$^m$54$^s$.219(15) &17$^o$46$'$13\farcs56(22)&\\
&PSR B0656+14  && 06$^h$59$^m$48$\fss$1503(7)  &14$^o$14$'$21$\farcs$1575(10)&
&06$^h$59$^m$48$^s$.152(14) &14$^o$14$'$21\farcs41(21)&\\
\hline
\end{tabular}
\label{t:coo}
\end{table*}
\begin{figure*}[tbh]
\vspace*{0.3cm}
\setlength{\unitlength}{1mm}
\resizebox{12cm}{!}{
 \begin{picture}(100,50)(0,0)
\put(0,0){\includegraphics[height=5.15cm,bb=102 174 545
633,clip]{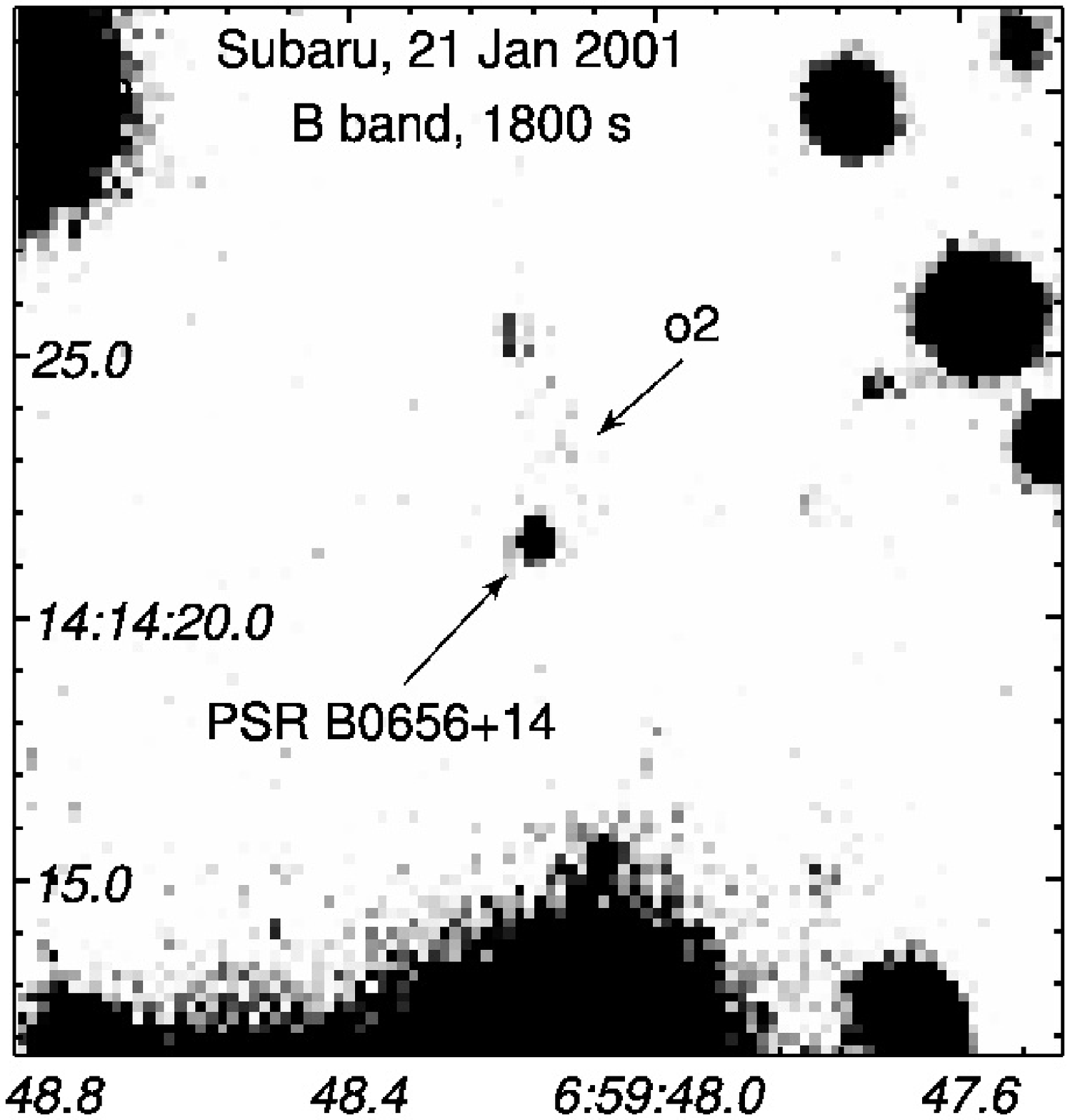}}
\put(50,0){\includegraphics[height=5.15cm,bb=96 174 545
633,clip]{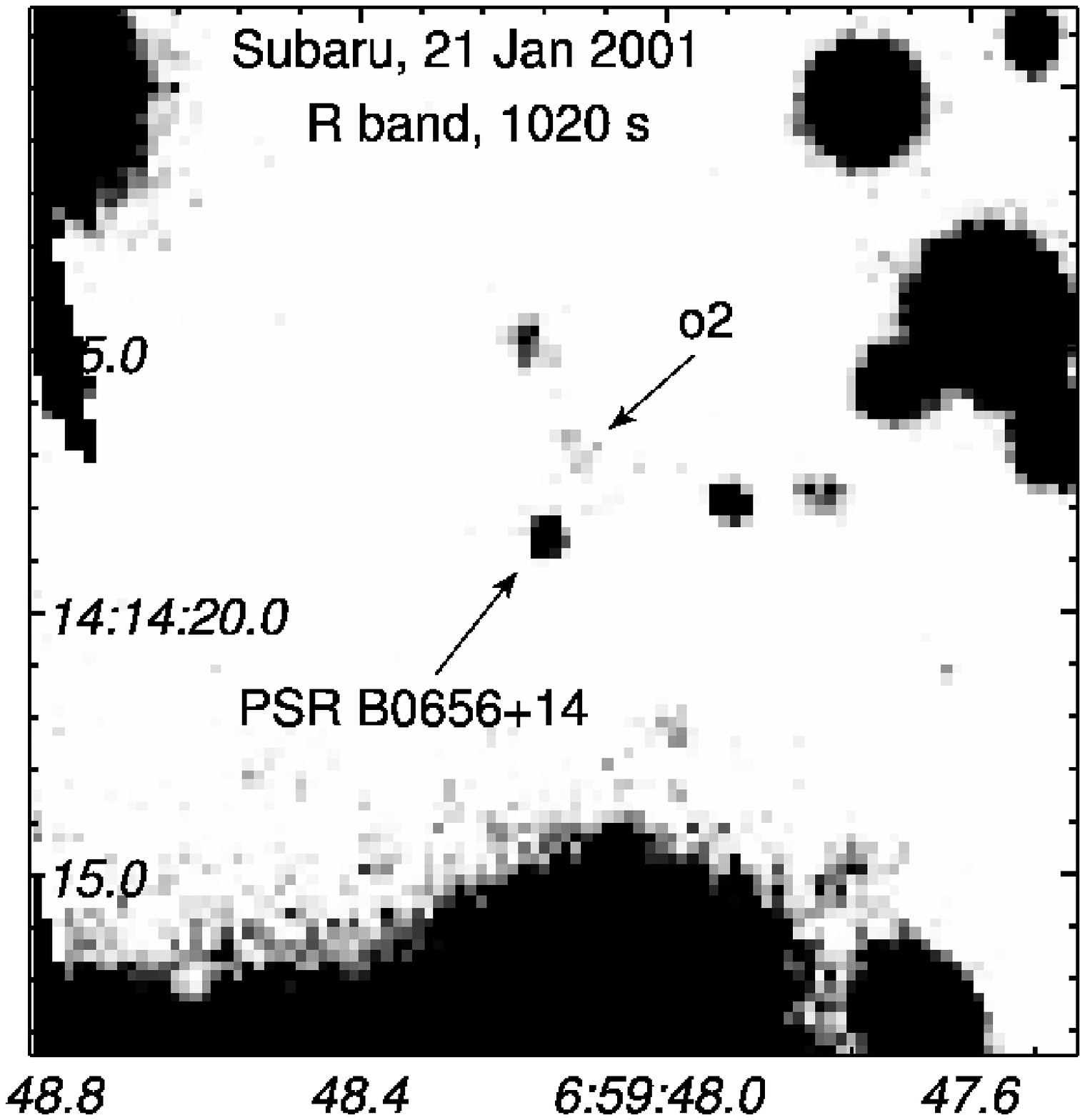}}
\put (100,0,){\includegraphics[height=5.15cm,bb=96 174
545 633,clip]{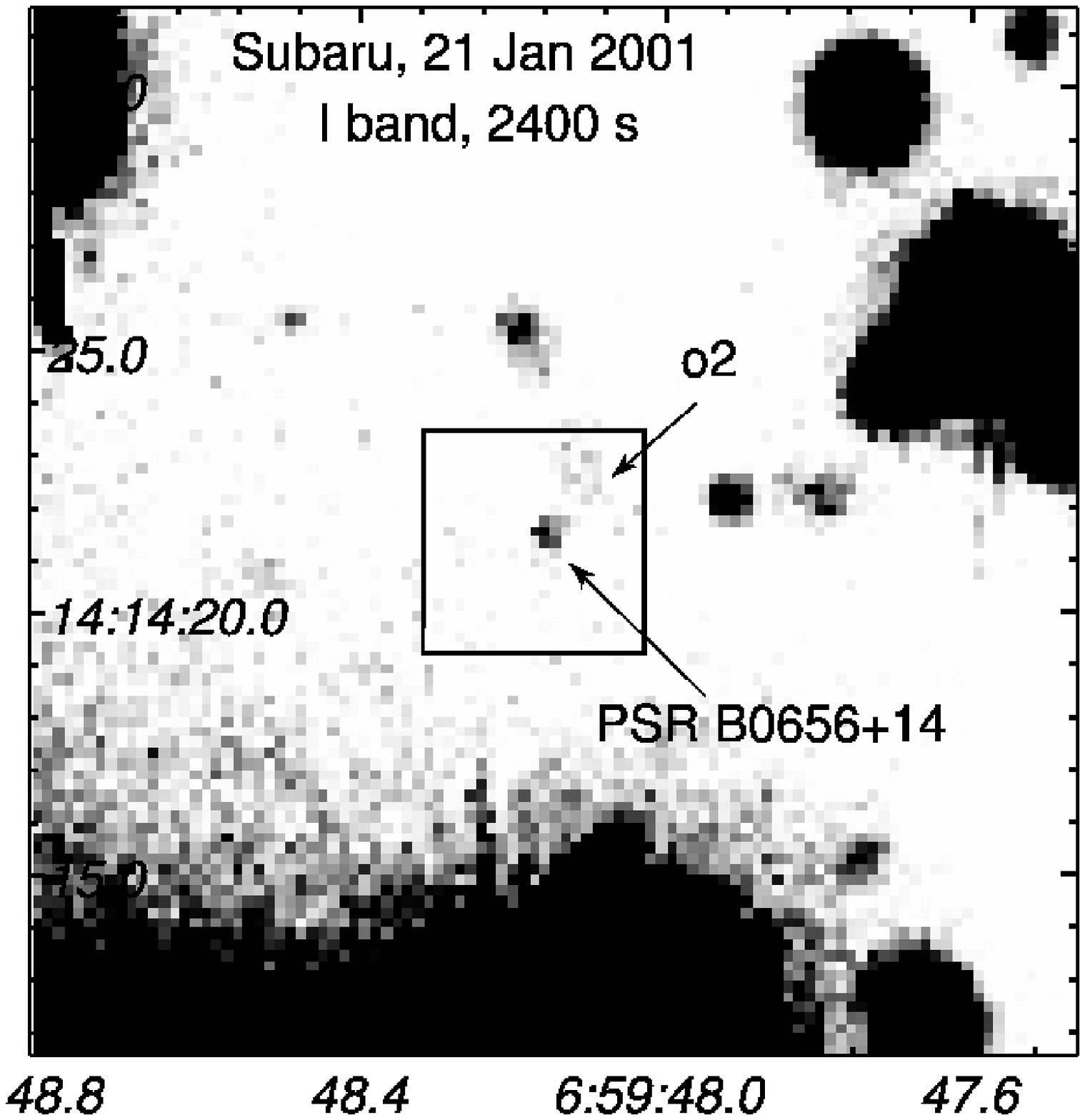}}
\end{picture}
}
\caption[h]{ 
$ 20\arcsec \times 20\arcsec$ fragments of the Subaru images 
of the \psh\ field in the $B$, $R$, and $I$ bands.
Units of RA (horizontal axis) are hours, minutes and seconds, and units
of Dec (vertical one) are degrees, arcminutes and arcseconds. A box in the {\it right panel} 
marks the region which is magnified in Fig.~\ref{f:656Imag}  and discussed
in~ Sect.~\ref{s:656res}.}
\label{f:656ima}
\end{figure*}

Standard data reduction  including bias subtraction and flat-fielding  
was performed making use of the {\tt MIDAS} and {\tt IRAF} software. 
In order to get rid of the remaining cosmic rays and to decrease possible
contamination of \psh\ by a nearby extended object we combined the individual
dither images of the pulsar field in the $B$ and $I$ bands making use of the {\tt ditherII}
package by Fruchter et al.~\cite{Fruchter}. 
\subsection{Astrometric referencing of the Subaru images} 
The positions and proper motions of \gem\ and \psh\ are well established (see Table~\ref{t:par}).
To determine the expected coordinates of \gem\ at the epoch of the Subaru 
observations (2001.054) its position  at the epoch of the
observations with HST/WFPC2 (1995.21) (Caraveo et al. \cite{Car96}) was used as
a reference point  and then corrected for the proper motion.
 The same was done for \psh\ using a reference point from
 Brisken et al.~(\cite{brisk03}).   based on the radio observations with the VLBA. The expected pulsar coordinates
are given in Table~\ref{t:coo} with errors accounting for  the proper motion
and reference point uncertainties.

The astrometric referencing of the Subaru images was done making use of 
the USNO-B1.0 catalog (Monet et al. \cite{Monet}) and  {\tt IRAF} tasks
{\tt ccmap/cctran}. For the astrometric transformation of  the combined images of
\begin{figure}[tbh] 
\begin{center}
\setlength{\unitlength}{1mm} 
\begin{picture}(54,108)(0,0)
\put(0,54){\includegraphics[width=5.1cm,bb=130 205 499
589,clip]{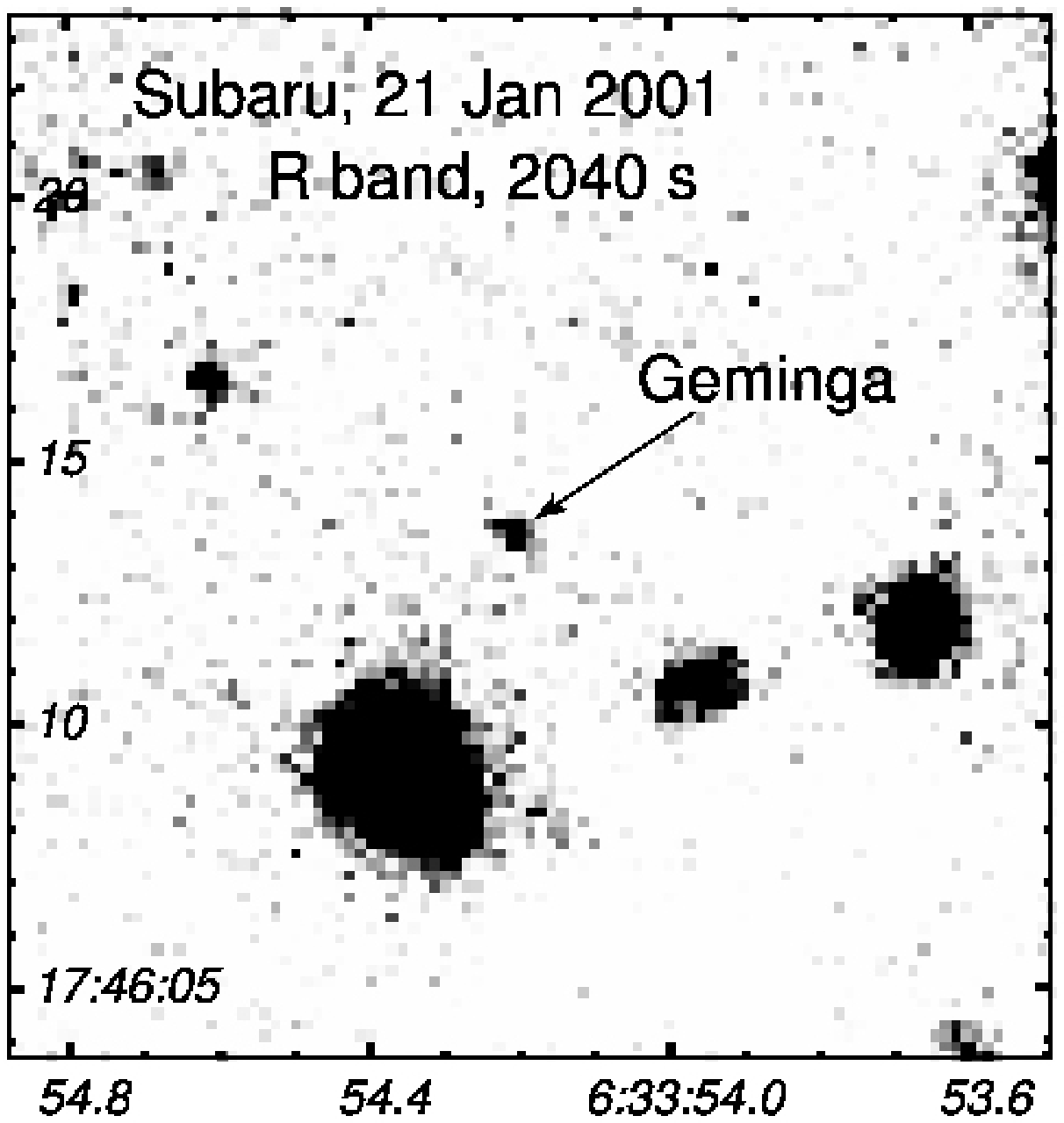}}%
\put(0,0){\includegraphics[width=5.1cm,bb=130 205 499
589,clip]{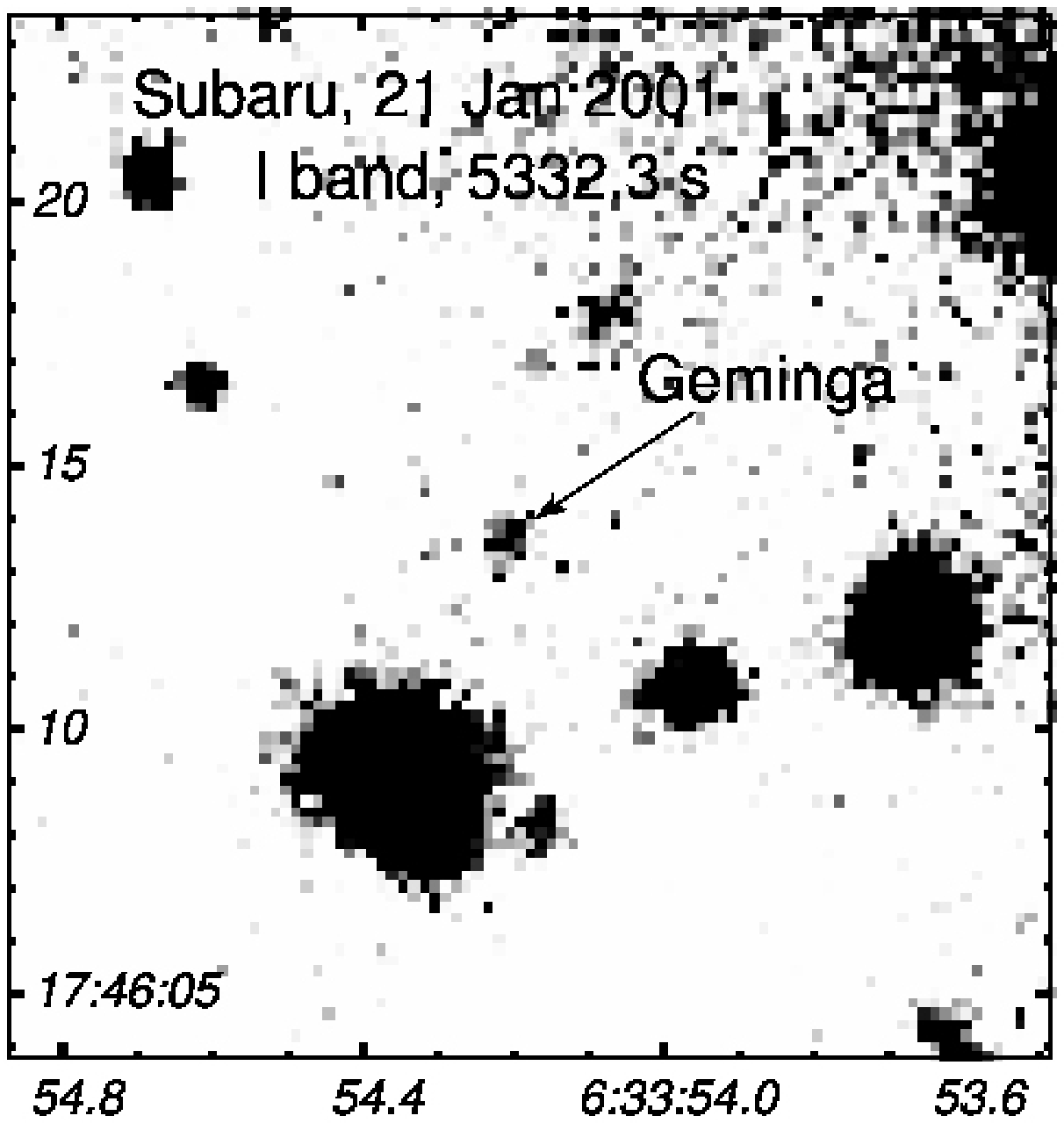}}
\end{picture} 
\end{center}
\caption[h]{$ 20'' \times 20''$ fragments of the Subaru images of the
\gem\  field in the $R$ ({\sl top})  and $I$ ({\sl bottom}) bands.  
Axes orientation and  label notations are the same as in Fig.~1.}  
\label{f:633ima}
\end{figure}
the \gem\ field  we used the positions
of six reference stars\footnote{USNO-B1.0 stars used for the  astrometric transformation of the \gem\ field: 
 1077-0137329, 1078-0139943, 1077-0137377, 1078-0139792, 1078-0139780 and 1077-0137166}. 
Formal {\it rms}  errors of the astrometric fit for the RA and Dec were 
$\approx0\farcs07$ and $\approx0\farcs06$, respectively, 
 for both the $R$ and $I$ images. 
Seven reference stars\footnote {USNO-B1.0 stars used  for
the   transformation of the \psh\ field: 
1042-0123510, 1042-0123514, 1042-0123581, 1042-0123600, 1042-0123481,
1042-0123424 and 1042-0123352} 
were used in case of \psh. For each of the $B$, $R$, and $I$ images formal {\it rms}  errors 
 were $\approx0\farcs08$  and $\approx0\farcs05$, for the RA and Dec, respectively,
which is less than the nominal USNO-B1.0 catalogue accuracy of
0\farcs20. Combining the {\it rms} errors and  the catalog uncertainty   
 we estimate the total 1$\sigma$  accuracy of  our astrometric referencing  of all the images 
as 0\farcs21 in both RA and Dec. 
The optical counterparts to both pulsars are clearly seen at the expected
positions  (see Fig.~\ref{f:656ima} and Fig.~\ref{f:633ima}).
The expected and observed positions (Table \ref{t:coo}) coincide with each other
within the errors accounting for the centroid and astrometric referencing  uncertainties. 
\subsection{Photometric referencing of the Subaru images}
Photometric calibration was carried out 
using three standard stars from the Landolt field PG1047+003 
(Landolt \cite{lan92}), observed the same night 
 and five   secondary standard stars from the \psh\ field (Kurt et al.~\cite{Kurt98}).
The atmospheric  extinction factors  in the  $B$, $R$, and $I$ bands
were derived from the variation of the count rates of field stars  with the airmass
changing during our observations  
(see Table~\ref{t:log}). The values obtained $k_B=0$\fm$18\pm 0.02$,
 $k_R=0$\fm$117\pm 0.021$, $k_I= 0$\fm$042\pm 0.020$
are consistent with the mean values at the Subaru site (Y.~Komiyama, private communication). 
Insignificant decrease of the extinction, within  1$\sigma$ level, 
was noticed from the beginning to the end of observations.   
Based that the following transition 
equations related to the observational night were obtained:
\begin{equation}
\eqalign
{
B - b  =  28.283(20) + 0.094(10) \cdot (b - r)  \cr
R - r  =  28.122(29) - 0.023(14) \cdot (b - r)   \cr
R - r  = 28.070(44) - 0.138(47) \cdot (r - i)   \cr
I - i  =  27.340(38) - 0.152(55) \cdot (r - i)   \cr
}
\label{eq:night}
\end{equation} 
where $b$, $r$, $i$ are  instrumental magnitudes and $B$, $R$, $I$ are 
 magnitudes in the Johnson-Cousins  photometric system.
The errors of the transformation coefficients shown in
parentheses  include 
the  extinction factor and  standard instrumental magnitude uncertainties.  
\subsection{The data from the ESO/NTT and HST archives}
\begin{figure*}[t]
\setlength{\unitlength}{1mm}
\begin{picture}(180,60)(0,0)
\put(0,0){\includegraphics[width=5.7cm,
bb=122 195 489 571, clip]{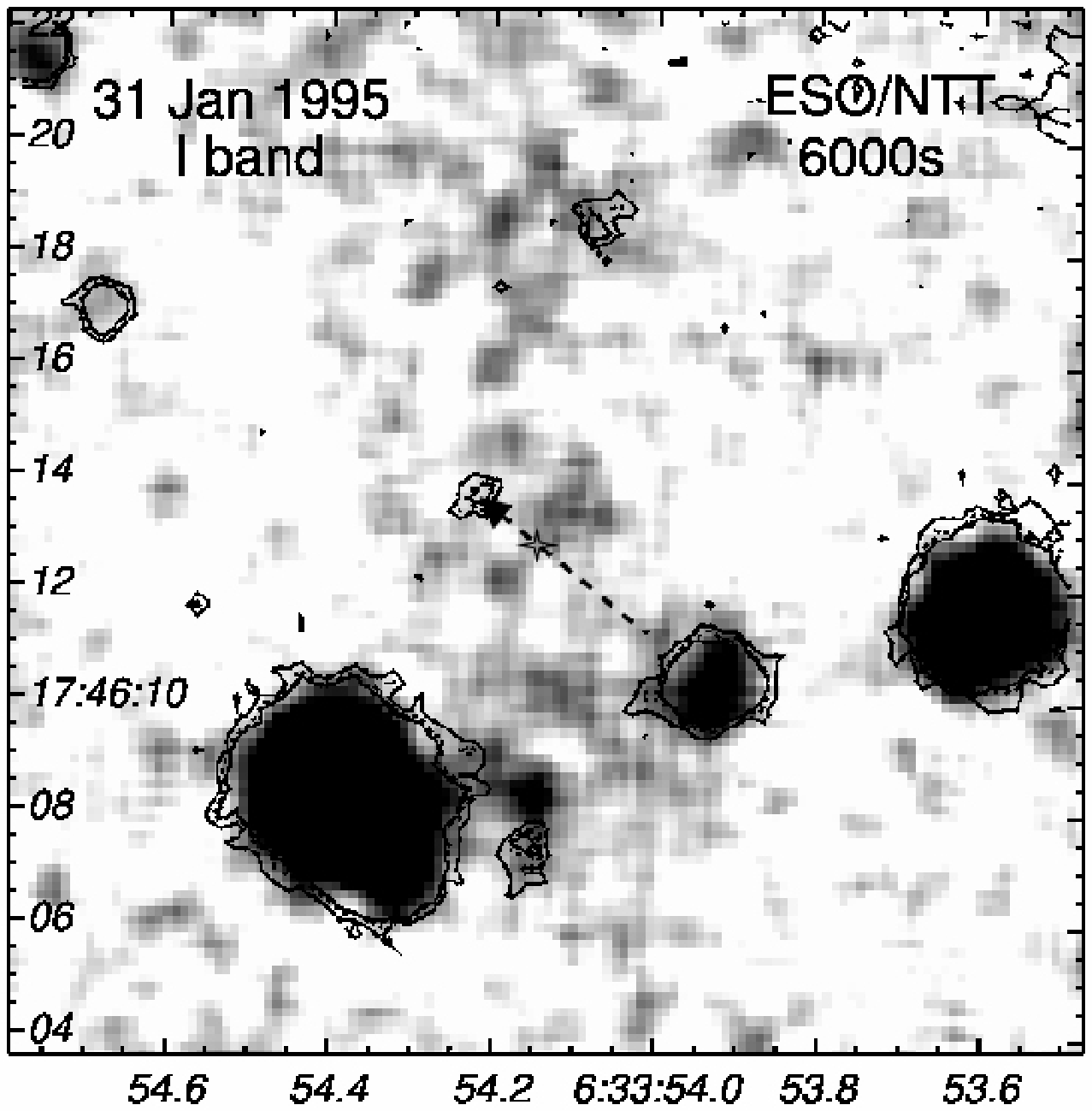}}
\put(60,0){\includegraphics[width=5.7cm,
bb=29 29 582 582,clip]{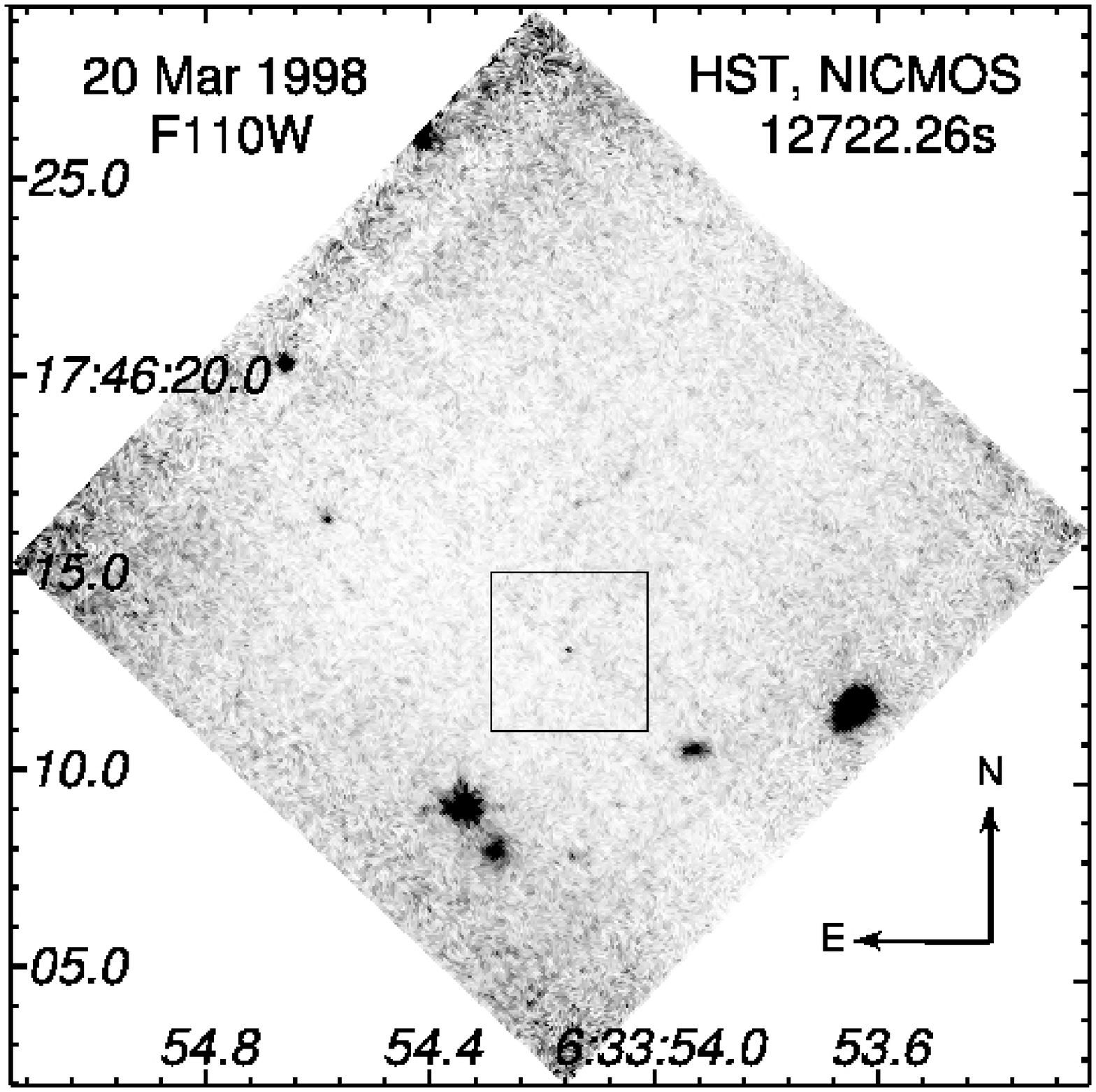}}
\put(120,0){\includegraphics[width=5.7cm,
bb=29 29 582 582,clip]{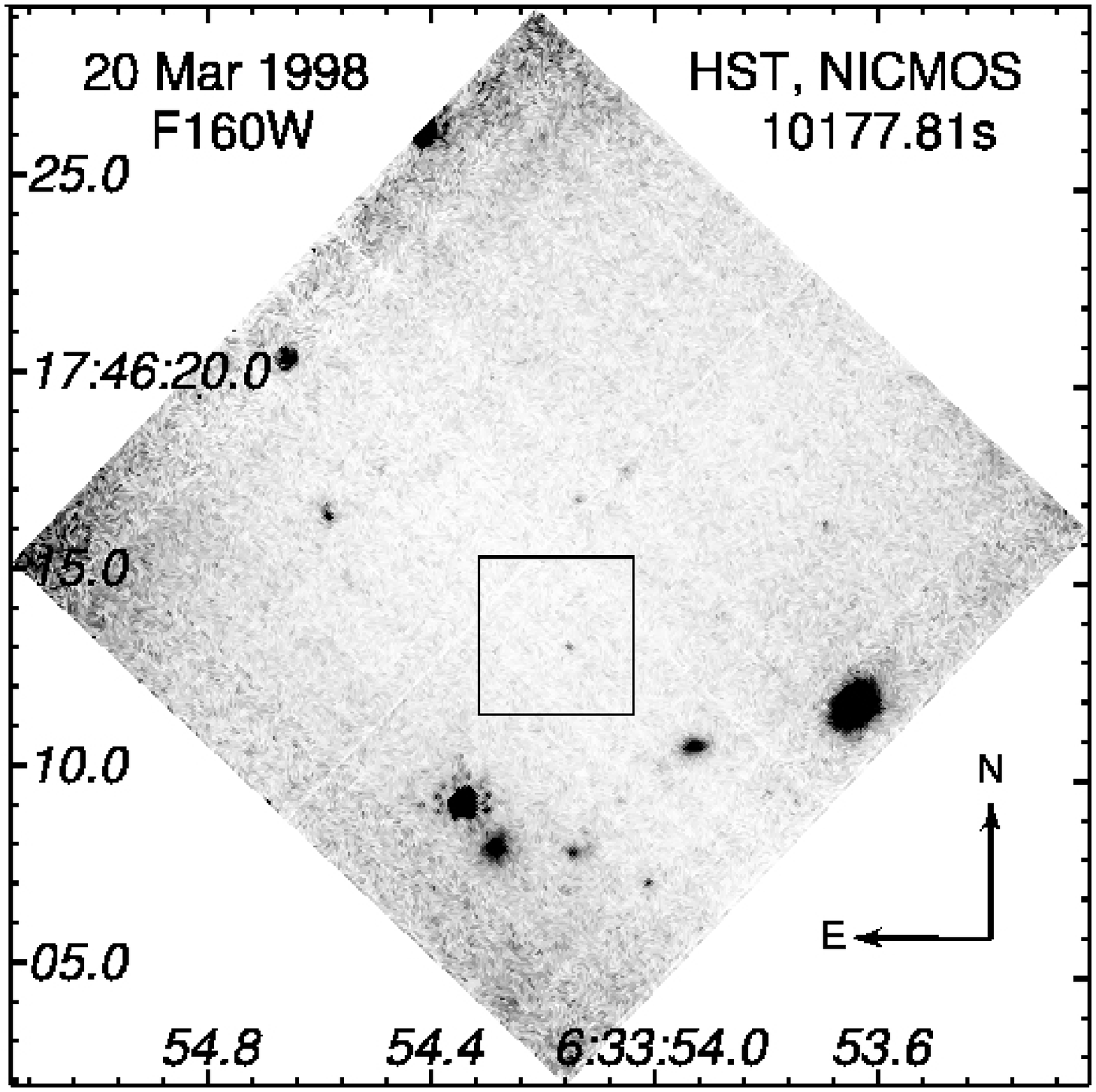}}
\end{picture}
\caption[h]{
{\sl Left:} A fragment of the \gem\ field image obtained 
with the ESO/NTT in the $I$ band (Bignami \etal \cite{big96}). 
The contours of the
Subaru $I$ band image are overlaid and the dashed-line arrow shows the
direction of the pulsar motion;  the cross marks the
position of the pulsar at the epoch of the ESO/NTT observations.    
{\sl Middle and right:} Near-infrared images  of the same field obtained
with the HST/NICMOS in the $F110W$ and $F160W$ bands, respectively (Harlow et al. \cite{Harlow}).
Boxes mark regions with Geminga,  they are magnified in Fig.~\ref{f:633imaH}. 
}
\label{f:633imaN}
\end{figure*}

In our analysis we used also the I-band images of the Geminga field obtained
with the ESO/NTT (Bignami et al.~\cite{bcmeb96}),
kindly provided by R.~Mignani. Five 20 min science
exposures were obtained on January 31 1995 using 
the SUSI\footnote{Observational conditions and other details can be found in Bignami et al.~\cite{bcmeb96}}.
 The reduction of these data  was rather complicated  due to the absence 
 of a standard  ``flatfield''  image  in the archive.    
 To produce an artificial flatfield  we subtracted stars from each 
 science exposure and combined  these images into   a median one.
 It was slightly improved in the regions around
 the star positions  making use of the {\tt MODIFY/AREA} task of the {\tt ESO/MIDAS} package,
 smoothed with the $5\times5$ pixel average and used as the  flatfield  
 in a standard data reduction.   The problems with the flatfield and the fact that 
 the individual science images were taken without shifts enough 
 to remove accurately all artifacts led to the appearance of     
artificial extended  objects around the position of Geminga 
 in the final image (Fig.~\ref{f:633imaN}, {\sl left panel}), 
 precluding  from a reliable identification of the pulsar (see Sect. 3.1. for details).  
\begin{table*}[tbh]
\caption{Flux measurements of the  pulsar optical counterparts (see 
 Sect.~2.3 for details).}
\label{t:fluxes}
\hspace*{0.5cm}
\vbox{
\begin{tabular}{l|ccccc|cl|ll}
\hline\hline
\multicolumn{1}{c|}{Object} &Band   &seeing$^a$  &$S/N$& $d^b$&Source$^c$ &  $k/\cos(Z)^d$ & $\delta_{\rm fin}d^e$& Mag$^f$     & Flux \widerul  \\
	      & &pix. &           & pix    & -2.5~Log(counts/s)  &  mag  & mag&   & $\mu$Jy \\
\hline
PSR B0656+14 &$B$& 4.5 & 22  &    4 &  -1.68   & 0.25    &     1.12       & 25.27(7)  &  0.313(20) \widerul \\
             &$R$& 3.4 & 18  &    4 &  -2.50   & 0.07    &     0.89       & 24.65(8)  & 0.417(30)           \\
	     &$I$&4.1  & 11  &    4 &  -1.90   & 0.05    &     0.95       & 24.52(13) &   0.371(42)\\
\hline
Geminga&$R$&3.7 & 15  &     4  &-1.63   & 0.14  & 0.84   & 25.53(11) & 0.185(17) \\
              &$I$&3.9 & 10  &     4  &-1.37   & 0.05  & 0.87   & 25.10(14) & 0.217(26) \\
\hline
\end{tabular}
}{
\renewcommand{\arraystretch}{0.8}
\hspace*{1cm}
\begin{tabular}{lll}
$^a$~FWHM of the stellar profile in the combined images.   &&  $^d$~Corrections for the atmospheric extinction. \\
$^b$~Diameter of the optimal photometrical aperture.  && $^e$~Correction factors for the PSF.  \\
$^c$~Instrumental magnitudes, $m = Source -  k/\cos(Z)  -\delta_{\rm fin}d$. && $^f$~Johnson-Cousins  magnitudes. \\
\end{tabular}
}
\end{table*} 

The NIR-optical-UV data from the HST archive\footnote{\href{http://archive.stsci.edu}{archive.stsci.edu}},
obtained with the HST/NICMOS  in  the $F110W$ and $F160W$ bands
(HST analogs of  $J$ and $H$;  Harlow et al.~\cite{Harlow}) and
with the HST/WFPC2  in the $F555W$ band for Geminga, with the HST/FOC
 in the $F195W$,  $F342W$,  $F430W$ bands for Geminga  (Bignami et al.~\cite{bcmeb96}; Mignani et al.~\cite{mcb98}) 
and  for  \psh\  (Pavlov et al. \cite{PSC}; \cite{PWC}) were also reanalysed.
The data reduction  and analysis of the  NICMOS images  obtained  at the epoch of 1998.22  
were performed as described in K\cite{K2001}\footnote{See also the HST Data Handbook for
NICMOS, \href{http://www.stsci.edu/hst/nicmos/documents/handbooks/}
{stsci.edu/hst/nicmos/documents/handbooks/}}.  
The photometry of the WFPC2 images processed by pipeline was performed using the algorithm
described in the WFPC2 Data analysis
Tutorial\footnote{Baggett et al. 2002, in HST WFPC2 Data
Handbook, v4, ed. B. Mobasher, Baltimore, STScI}. 
The reduction of the FOC images was done in the same  manner as described in 
Pavlov et al.~(\cite{PSC} and \cite{PWC}) making use of  the FOC
Instrument Handbook (v6)\footnote{\href{http://stsci.edu/ftp/instrument
news/FOC/Foc handbook/foc handbookv6.html}
{stsci.edu/ftp/instrument news/FOC/Foc handbook}}. 
\section{Results} 
\label{s:res}
The optical counterparts of both pulsars  are firmly detected in all  
Subaru images  with a signal to noise  ratio $S/N\ge 10$. 
The instrumental magnitudes of the  counterparts were
measured for a range  of circular aperture radii of
($1-3$)~CCD pixels centered at the source positions. They were corrected for 
the point spread function (PSF) of bright  stars. The background was 
measured over the annulus with the inner and outer radii $r_{in}=5$ and $r_{out} = 10$ pixels.
Within the measurement errors the results for different  apertures coincided
and a 2 pixel radius, or 4 pixels in the diameter that is comparable with
the seeing of about 3-4 pixels (cf.~Table~2), was  adopted as an optimal
to get a maximum  $S/N$. Using  Eq (1) Johnson-Cousins magnitudes  
were derived and transformed into the fluxes in physical units  
using  zero-points  provided by Fukugita et al.~\cite{Fukugita}.  
The results are given in Table~\ref{t:fluxes}. 
The errors include  statistical errors of the instrumental magnitude measurements, the errors of
photometric referencing, and possible atmospheric extinction variations. 
\subsection{Geminga}
\subsubsection{Identification in the I band}
The object of $I$$\approx$25\fm1 detected with the Subaru 
at the expected  Geminga position  is by a magnitude  brighter  than the upper 
limit $I$$\ge$26\fm4 reported by Mignani et al.~(\cite{mcb98}) 
based on  previous  $I$ band observations  with the ESO/NTT.  
Such a deep  limit is puzzling since it is comparable with the 3$\sigma$  detection limit 
of our Subaru observations, I$\ge26\fm8$, obtained with  similar exposure time  
(1.5\,h {\it vs} 1.4\,h of NTT), but with a much higher telescope collective area  (8\,m  {\it vs}
3.6\,m of NTT) and at  better seeing conditions  (0\farcs6--0\farcs8 {\it vs} 0\farcs9--1\farcs1 of NTT).  
Using the NTT/SUSI {\it Exposure Time Calculator} (ETC)\footnote{eso.org/observing/etc/}         
one can easily estimate that at the respective conditions the NTT 
image should be by a magnitude less deep than the reported limit. 
This is in  agreement with an upper limit of $I$$\ge$25\fm4, which is   
presented in the text of the paper by Bignami et al.~(\cite{bcmeb96}) 
describing  the NTT observations of Geminga. However, this reasonable limit was  likely  incorrectly 
reproduced  in  Figs.~3 and 4 of the same paper, and  later on in Mignani et al.~(\cite{mcb98}). 
  
\begin{figure}[tbh]
\setlength{\unitlength}{1mm}
\begin{picture}(90,45)(0,0)
\put(0,0){\includegraphics[width=4.3cm, 
bb=29 29 582 587,clip]{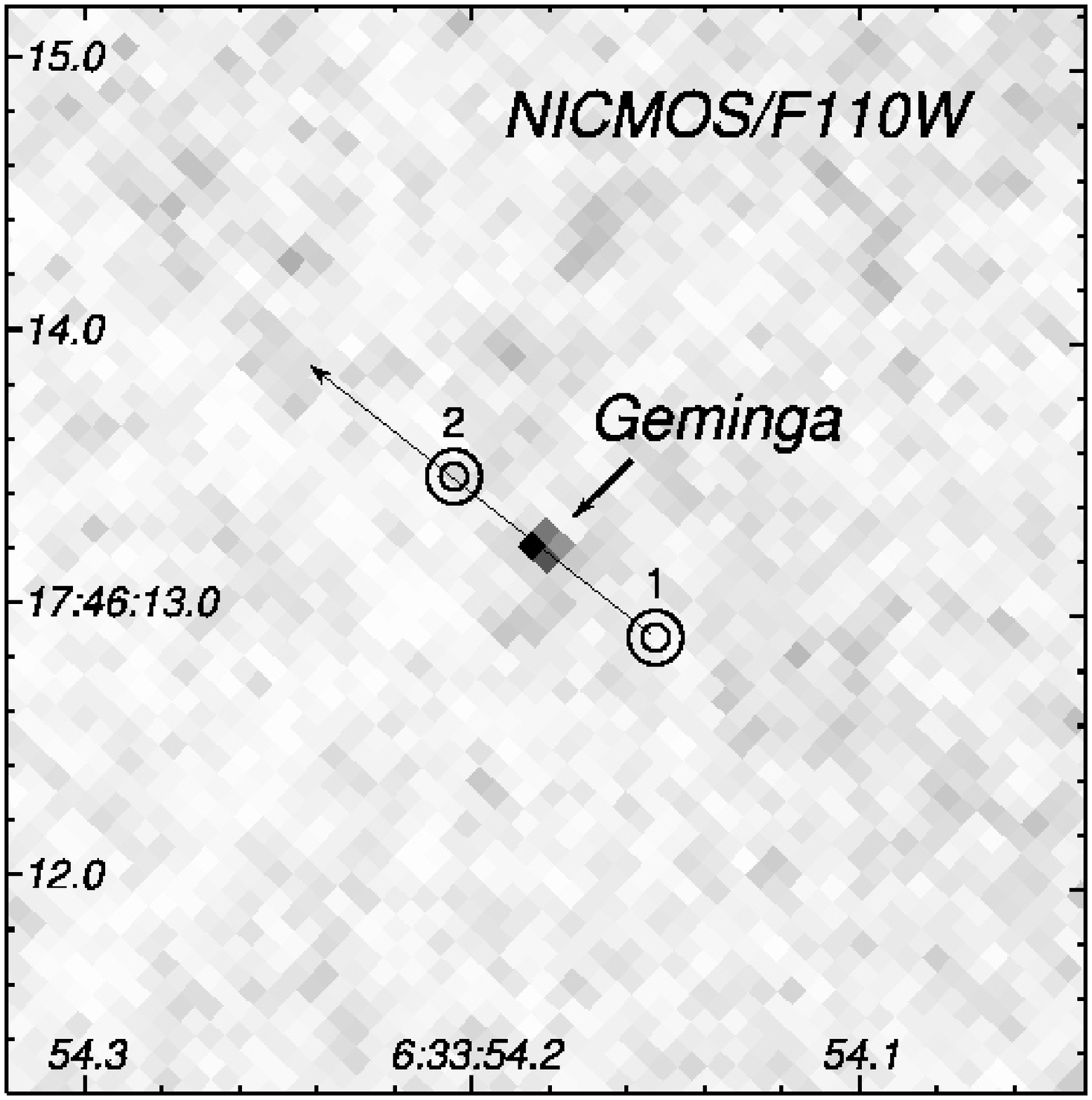}}
\put(45,3){\includegraphics[width=4.3cm, 
bb=29 29 582 582,clip]{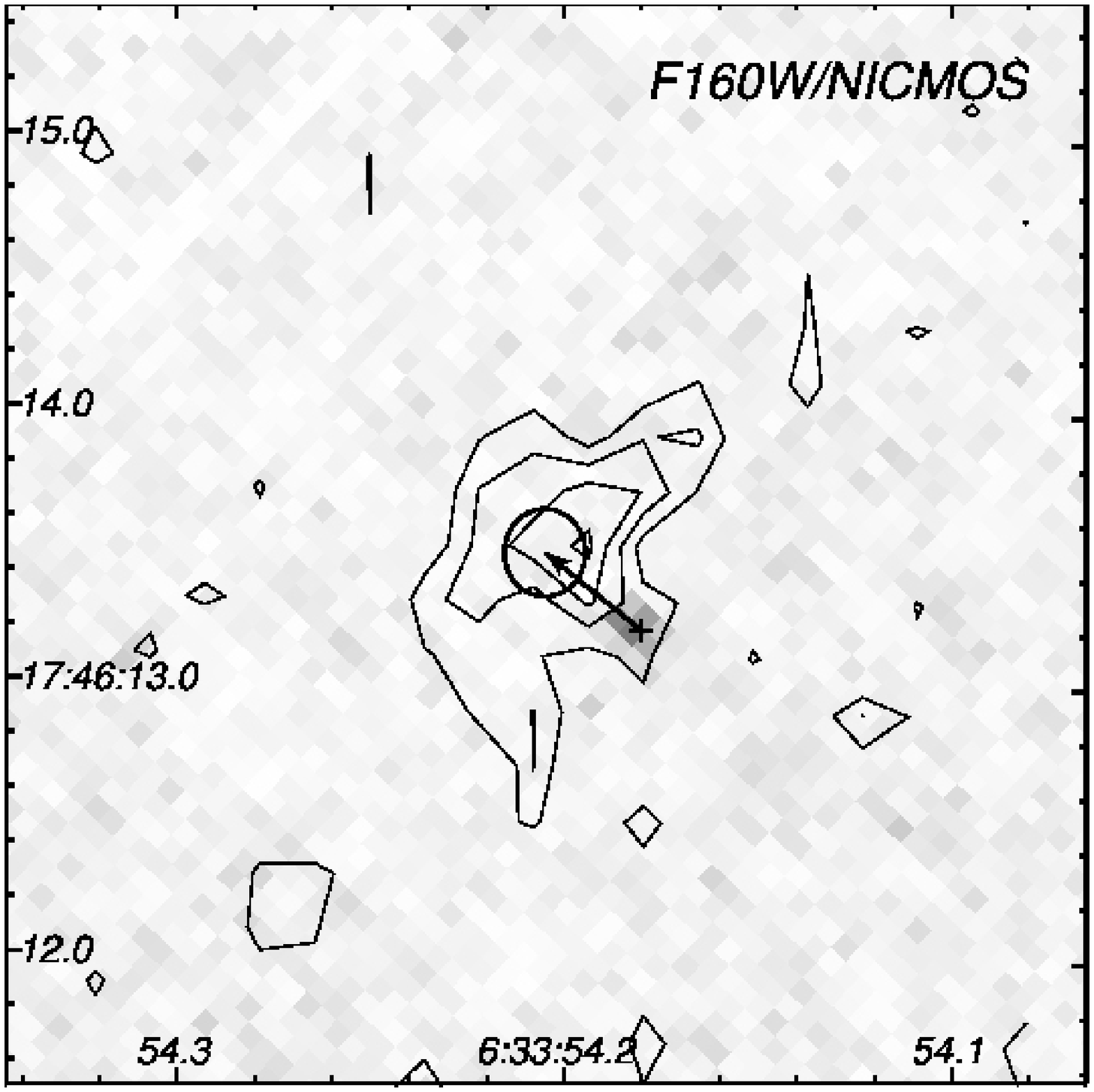}}
\end{picture}
\caption[h]{  $4'' \times 4''$ images of Geminga  in the $F110W$ 
({\sl left}) and  $F160W$ ({\sl right}) bands.  
The circle annulus in the {\sl left panel} mark the expected
Geminga positions at the epochs of I-band observations with 
the NTT (1) and Subaru (2), respectively, while the long arrow shows 
the pulsar path calculated based on the proper
motion measurements of Caraveo \etal~(\cite{Car96}). 
Contours of the Subaru image in $I$ band are overlaid in the  
{\sl right panel} where  the arrow  shows the shift of \gem\ due to its proper motion over 3.2~yr 
between the HST and Subaru observations;  ``$+$'' and 1$\sigma$-ellipse mark 
the expected positions of the pulsar 
at the epochs of the HST and Subaru observations, respectively.}
\label{f:633imaH}
\end{figure}

 To solve this puzzle  we  inspected  the archival NTT image, which is shown 
in the { \it left panel} of Fig.~\ref{f:633imaN}. 
The expected position of Geminga at the epoch of the NTT
observations,  1995.084,  is marked by a cross and the proper motion of the pulsar is shown by a dashed arrow.    
Several background-like objects are seen in this image in a few arcseconds 
from the position of Geminga, while the pulsar itself  can be hardly identified in this image. 
A thorough comparison with the Subaru $I$-band  image  ({\sl
bottom panel} of Fig.~\ref{f:633ima}), whose contours are overlaid,  
shows  that these objects are  likely to be artifacts caused by a poor data 
reduction in the absence of the standard flatfield image in the archive (cf.~Sect.~2.4).  
This is further strengthened by the fact that there are no
background objects in this region detected neither in our
Subaru  $R$ band   ({\sl top panel} of Fig.~\ref{f:633ima}) nor in the HST $F110W$ band images  
(Fig.~\ref{f:633imaN},~\ref{f:633imaH}) which are adjacent to $I$ band  
in the spectral energy distribution and where Geminga 
is clearly identified. Our rough estimate of the 3$\sigma$  
detection limit of the  NTT image shown in  Fig.~\ref{f:633imaN}     
gives $I$$\ge 24\fm4$. This is by a magnitude 
less deep than the above ETC and those of  Bignami et ai.~(\cite{bcmeb96})  estimates  
and  can be caused by the data reduction
problems mentioned above.  However,
it is hardly possible to get  a two magnitude deeper flux
limit, as  in Mignani et al.~(\cite{mcb98}),   
by any improvement of the  NTT data reduction. Therefore,  
it has been overestimated at least by a magnitude for unknown reasons.
On the other hand, we note   that the object detected at the Geminga position in the Subaru $I$ image   
could be likely detected also in the NTT image with
$S/N\sim 4$, provided a straightforward NTT data reduction is possible.  
This  would be consistent with the significance of a
marginal  detection of Geminga in the $I$ band with the 6m BTA telescope 
reported by Kurt et al.~(\cite{Kurt2001}).       

To  insure further our  $I$ band  and NIR identification of Geminga  
we estimated the shift between the counterpart positions 
in the HST/NICMOS and Subaru images (see Fig.~\ref{f:633imaH}). 
It was found to be 0\farcs45$\pm$0\farcs22. Despite the only 
$2\sigma$ significance of this estimate  it is in a good agreement 
with the shift direction (marked by arrows in Fig.~\ref{f:633imaH}) 
and with its value of  0\farcs31$\pm$0\farcs01,  
which are expected from the Geminga proper motion 
(Caraveo et al. \cite{Car96}; Table~\ref{t:par}) 
at the time base of about 3.2~yr separating the HST 
and Subaru observations.   We have not found also    
any reliable faint red background source on the path of the pulsar proper motion, 
which would prevent from  correct identification of Geminga in the $I$ band  
at epochs after the NTT observations, as was warned by R.~Mignani 
(private communication). This allows us to make a conclusion 
on a firm identification of Geminga  in the $I$, as
well as in $R$, and  in
the two NIR bands. 
\subsubsection{Broadband fluxes from the NIR through FUV}  
The results of our I-R and NIR photometry of Geminga  are presented in
Table~\ref{t:633_fl} where we have collected also  most
\begin{table}[tbh]
\caption{Geminga  in the NIR-optical-NUV-FUV range.
}
\label{t:633_fl}
\begin{tabular}{lccc}
\hline\hline
Filter/  & Log ${\nu}$        & $F_{\nu}$   & $F_{\nu}$  \widerul \\
 Telescope                         &              &
 observed         &dereddened$^a$   \\
                   &  Hz            & $\mu$Jy          &$\mu$Jy    \\
\hline
\multicolumn{4}{c}{NIR} \widerul\\
\hline
$F160W$/HST          & 14.273($^{+58}_{-51}$)          &
0.230(25)$^{b,c}$      &0.233(25)   \\
$F110W$/HST          & 14.436($^{+138}_{-105}$)           &
0.166(21)$^{b,c}$      &0.170(21)   \\
\hline
\multicolumn{4}{c}{optical} \widerul\\
\hline
$I$/Subaru           & 14.571($^{+44}_{-40}$)    &
0.193$^{+19}_{-21}$$^c$&0.201$^{+20}_{-22}$   \\
$R$/Subaru           & 14.658($^{+55}_{-49}$)    &
0.172$^{+17}_{-18}$$^c$&0.181$^{+18}_{-19}$   \\
$F555W$/HST          & 14.734($^{+62}_{-54}$)           &
0.170(20)$^d$            &0.182(21)   \\
$F555W$/HST          & 14.734($^{+62}_{-54}$)           &
0.146(13)$^c$          &0.156(14)   \\
$F430W$/HST          &14.882($^{+49}_{-44}$)            &
0.146(11)$^c$          &0.160(12)   \widerul\\
\hline
\multicolumn{4}{c}{NUV} \widerul\\
\hline
$F342W$/HST          & 14.944($^{+47}_{-43}$)           &
0.201(10)$^c$          &0.224(11)   \\
$F25SRF$/HST       & 15.116($^{+122}_{-95}$)         
&0.261(26)$^d$     &  0.315(31) \\
F195W/HST          & 15.104($^{+159}_{-39}$)           &
0.288(30)$^c$          &0.344(36)  \\
\hline
\multicolumn{4}{c}{FUV} \widerul\\
\hline
$G140L$/HST         & 15.322(84)            &0.455(29)$^d$  
     & 0.546(35) \\      
\hline
\end{tabular}
\renewcommand{\arraystretch}{0.8}
\begin{tabular}{ll}
$^a$~E(B-V)=0.023;  &$^c$~this work;  \\  
$^b$~Harlow \etal~\cite{Harlow};  &$^d$~Kargaltsev \etal~\cite{Karg05} using ACS  \\
&  and STIS/MAMA NUV and FUV. \\
\end{tabular}
\end{table}
 accurate and partially updated results of available pulsar 
observations in other broad bands.     The flux  in the $R$ band  
is in a good agreement with what was published by Mignani et al.~(\cite{mcb98}),    
but the accuracy of the Subaru observations is  considerably 
higher.   The result in the $I$ band  confirms  a  tentative 
detection reported by Kurt et al.~(\cite{Kurt2001}),  
but at much higher significance level.  
The NIR fluxes are in agreement with the preliminary results
published by Koptsevich \etal~(\cite{K2001}) 
and Komarova \etal~(\cite{K2003}).   

Our reanalysis of  the four  HST/WFPC2 
datasets obtained  at different epochs  in the $F555W$ band
 in the period of 1994--1995   yield  the  Geminga fluxes  in the range of  0.136--0.161 $\mu$Jy
depending on the set.  The flux values in
separate sets are consistent  with each other and  do not show 
any  variation of the Geminga emission  with time above 3$\sigma$ level. 
In Table~\ref{t:633_fl} we show a mean $F555W$ flux  over these sets. 
It is also consistent  with the value obtained  recently   with the HST/ACS in
the same band (Kargaltsev et al. \cite{Karg05};
cf.~Table~\ref{t:633_fl}) but a  factor of 1.7 smaller
than early results  (Bignami et al.~\cite{big96}; Mignani et
al.~\cite{mcb98}).  This  provides independent confirmations  of a
significant overestimation  of  the Geminga flux in the
$F555W$ band  in the previous publications, 
as it has been noticed  by Kargaltsev et al.~(\cite{Karg05}).  
At the same time,  our reanalysis of the  HST/FOC 
archival data obtained in $F430W$,
$F342W$, and $F195W$ bands shows a less significant, about
2$\sigma$ (or 10\%--20\% of the flux value),  
but systematic underestimation of the previously published
fluxes.  In Table~\ref{t:633_fl}
we present the updated  flux values in these bands.            
\subsubsection{Possible bow-shock structure}   
Observations of the Geminga field in X-rays with the XMM have revealed 
a faint bow-shock structure  around this pulsar whose tails 
are extended  up to 2\farcm5 behind the  moving NS (Caraveo et al.~\cite{Car03}). 
Bow-shock structures  produced  by  interaction of 
supersonically moving NSs with
ambient medium have been detected around several pulsars
presumably in 
the H$_{\alpha}$ emission (e.g., the Guitar nebula around  PSR
B2224+65; Cordes et al.~\cite{cord93}), where they are usually
easier detected than in X-rays  (e.g., Romani et al.~\cite{romani97}).  
However,  no signs of any $H$$_{\alpha}$ counterpart of the  
Geminga shock structure  were detected
at  deep imaging  carried out in  narrow  $H$$_{\alpha}$
bands with the VLT  (Caraveo et al.~\cite{Car03}) and 6m BTA   
(Komarova et al.~\cite{K2005}). A faint  $H$$_{\alpha}$ counterpart,  if 
exists,  is  likely hidden  in a rather strong extended
background    
emission  in the $H$$_{\alpha}$ line from a low Galactic
latitude  region where 
Geminga is located  (b$\sim$4\fdg7). 
\begin{figure}[tbh]
\begin{center} 
\includegraphics[width=5.5cm,
bb=58 150 590 681, clip]{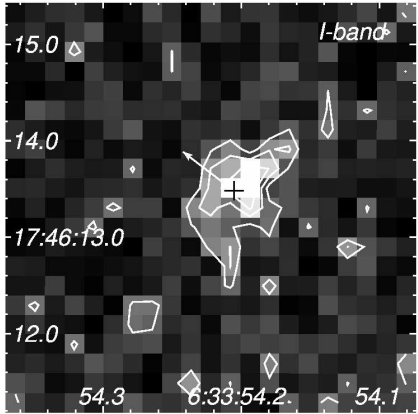}
\end{center} 
\caption[h]{ A magnified fragment of the Subaru $I$ band image of   
 Geminga  demonstrating an arc-like structure    
 resembling a head of a bow-shock 
 due to   the pulsar  supersonic 
motion in the interstellar matter (cf.~contours in
Fig.\ref{f:633imaH}). The motion direction is shown by
arrow, the pulsar position is marked by a cross.}
\label{f:633Imag}
\end{figure}    

At the same
time, as in case of the Crab pulsar nebula  (Hester et
al.~\cite{hes02}),   the brightness of the shocked 
structures of Geminga could be higher  in the optical
continuum instead of lines.  
In this case they have to be more easier detected     
with broadband observations.  
A careful inspection of the Subaru images has allowed us  to find  
possible indications of the optical counterpart of the
X-ray structure in  the  deepest $I$ band image. 
A faint  arc-like feature  resembling    
a head of the bow-shock can be seen in Fig.~\ref{f:633Imag}. 
Its contours are shown  in the  {\sl right panel} of Fig.~\ref{f:633imaH}. 
The  arc length is  $\sim 1\farcs9$  and it extends perpendicular 
to the pulsar path. There are no signs of this structure 
in our less deep $R$ band image as well as in the archival HST images.
Further deep imaging  of the field is needed to understand whether the 
arc-like structure is indeed a counterpart of the bow-shock seen in X-rays, 
a faint background extended object, or an artifact.        
\subsection{PSR B0656+14} 
\label{s:656res} 
\subsubsection{Broadband fluxes from the NIR through NUV} 
\begin{table}[tbh]
\caption{PSR~B0656+14  in the NIR-optical-NUV range}
\label{t:656_fl}
\begin{tabular}{lccc}
\hline 
\hline 
Filter/  & Log ${\nu}$        & $F_{\nu}$   & $F_{\nu}$  \widerul \\
 Telescope                         &              &
 observed         &dereddened$^a$   \\
                   &  Hz            & $\mu$Jy          &$\mu$Jy    \\
\hline\multicolumn{4}{c}{NIR} \widerul\\
\hline
$F187W$/HST        &           14.204($^{+30}_{-28}$) &
0.779(97)$^b$ & 0.789(98)      \\
$F160W$/HST        & 14.273($^{+58}_{-51}$)            &
0.575(41)$^b$ & 0.584(42)      \\
$F110W$/HST        &   14.436($^{+138}_{-105}$)    &
0.336(30)$^b$ & 0.346(31)      \\
\hline\multicolumn{4}{c}{optical} \widerul\\
\hline
$I$/Subaru         &  14.571($^{+44}_{-40}$)   &
0.371($^{+44}_{-40}$)$^c$& 0.390($^{+46}_{-42}$)       \\
$R$/Subaru         &   14.658($^{+55}_{-49}$)  &
0.417($^{+55}_{-49}$)$^c$& 0.447($^{+59}_{-53}$)      \\
$F555W$/HST        & 14.734($^{+62}_{-54}$)    &
0.393(24)$^b$ &  0.428(26)          \\
$B$/Subaru         & 14.829($^{+52}_{-47}$) &
0.313($^{+52}_{-47}$)$^c$  & 0.350($^{+58}_{-53}$)  \widerul \\
$F430W$/HST        &  14.882($^{+49}_{-44}$)                
  & 0.264(29)$^d$ &  0.300(33)       \\
\hline
\multicolumn{4}{c}{NUV} \widerul\\
\hline
$F342W$/HST        &   14.944($^{+47}_{-43}$)          &0.307(27)$^d$ &  0.353(31)     \\
$GRISM$/HST  &  15.071($^{+170}_{-125}$)    & 
0.354(31)$^e$ &  0.429(38)     \\
$F195W$/HST        &  15.104($^{+159}_{-39}$)           
&0.349(43)$^d$ &  0.438(54)     \\
\hline
\end{tabular}
\renewcommand{\arraystretch}{0.8}
\begin{tabular}{ll}
$^a$~E(B-V)=0.03;   & $^d$~Pavlov et al.~\cite{PWC};  \\ 
 $^b$~K\cite{K2001};  & $^e$~Shibanov et al.~\cite{Sh2005}.   \\  
$^c$~this work; &  \\  
\end{tabular}
\end{table}
Previous photometric measurements of the pulsar 
fluxes in the $B$, $R$, and,
especially in the $I$ band (Kurt et al.\cite{Kurt98}; K\cite{K2001}) were rather uncertain
due to poor seeing conditions and a contamination from a faint extended 
background object o2 which sits only in 1\farcs1 from the pulsar (cf. Fig.~\ref{f:656ima}). 
Perfect seeing during the Subaru observations allowed us to
avoid the contamination and to measure for the first
time the fluxes in these bands with the accuracy   of better than 10\%.
The results of our  photometry  of the pulsar 
are shown in Table~\ref{t:656_fl}, where  
the most accurate and informative  data  in other broad 
bands are also collected\footnote{As in case of Geminga (Sect.~3.1),  using archival data  we have 
examined  the published fluxes of PSR~B0656+14
obtained with different HST instruments. 
 No significant discrepancies from the
results as they are listed in K\cite{K2001} were found.}. 
\begin{figure}[tbh]
\begin{center} 
\includegraphics[width=5.5cm,bb=60 149 589
680,clip]{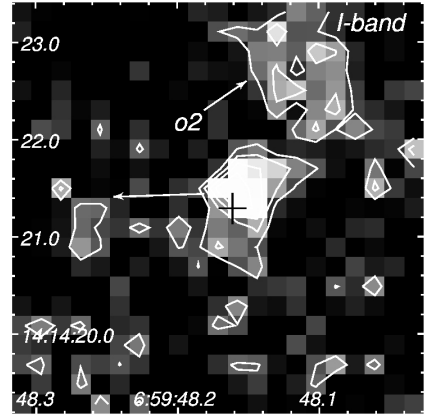}
\end{center}
\caption[h]{Magnified fragment of the Subaru $I$ band image of   
 \psh\   demonstrating possible existence 
 of a PWN or a bow-shock structure around the pulsar (extended object in the center of the image). 
 Contours 
 are overlaid to underline the shape of the structure;  
 the arrow shows the direction of the pulsar proper motion, the pulsar position is
 marked by a cross     
}
\label{f:656Imag}
\end{figure}    
\subsubsection{Possible PWN structure} 
As in case of Geminga we find also a hint of a faint compact nebula 
around  \psh\  possibly associated with the pulsar activity.   
This is seen in the deepest image obtained in the $I$ band (Fig.~\ref{f:656Imag}).
The detected structure is elongated approximately  perpendicularly to the direction of the pulsar 
proper motion.  Its  maximal  extension is $\sim 1\farcs5$.  
If this is not a background object or an artifact, it is the second 
indication of the presence of a faint compact PWN or a bow shock  nebula 
around PSR B0656+14 in the optical range. The first one was mentioned 
by Koptsevich et al. (\cite{K2001}) who found a knot-like structure 
seen at a 3$\sigma$ level above the wings of the pulsar spatial profile 
at $\sim 0\farcs04$ from the pulsar position in the archival HST/FOC/$F343W$ image. 
Knot-like structures are typical attributes of the Crab PWN.  
However, in case of \psh\  both indications still have to be considered 
with caution and need to be confirmed by deeper observations since 
we found no signs of the nebula in any other available 
optical images of the pulsar. However, all  these images are   
less deep than the Subaru $I$ band image.       
\section{Discussion}
\label{s:dis} 
Reliable detection of  \gem\ and \psh\  in the $I$ and $R$ bands  
basically completes a comprehensive broadband photometric study of both pulsars 
in the NIR through  FUV range started about a dozen years
ago. As a result,  the two pulsars  have been detected  at
a high significance level  practically in all standard
optical bands generally used for the  photometry  of ordinary stars to obtain  the spectral energy 
distributions of their radiation  in a wide spectral domain. 
Except for the most bright and younger Crab and Vela
pulsars,      \gem\ and \psh\ stand up to be  the most studied  up to
date  radio pulsars in the optical range.     
\subsection{Spectral energy distribution in the
optical range} 
\label{s:sed}
\subsubsection{Extinction corrections}
Interstellar color excesses  $E(B-V)$ along
the lines of sight to \gem\ and \psh\  have not yet been investigated accurately.
This requires additional careful optical studies  of suitable stars from 
the pulsar neighborhoods. So far there are only indirect estimations of
the extinction  based on observations in other spectral domains (e.g.,
Pavlov \etal~\cite{PSC}; Kurt \etal~\cite{Kurt98}; Kurt \etal~\cite{Kurt2001}). 
Closeness of both 
pulsars to the Galaxy plane (see Table~\ref{t:par}) suggests that their
optical fluxes  are to be affected by the foreground  extinction from
the Milky Way. On the other hand, the reddening cannot be  too strong due
to their proximity to the Earth. This is confirmed by  relatively low  DMs
and  hydrogen column densities  $N_H$  obtained from the radio and X-ray
observations.  The conservative $N_H$ values are
$\approx$$1.1\times 10^{20}$cm$^{-2}$ for \gem\ 
based on the ROSAT and ASCA X-ray spectral observations 
(Halpern \& Wang~\cite{hal97}; Jackson et al.~\cite{jak02}) 
and  $\approx$$1.4\times 10^{20}$cm$^{-2}$  for \psh\ based
on the ROSAT, ASCA and EUVE observations  (K2001).  
Using these values and an empirical relation $N_H/E(B-V) =  4.8\times 10^{21}$cm$^{-2}$mag$^{-1}$
applicable for the Milky Way (Bohlin \etal~\cite{boh1978}) one can obtain
$E(B-V)$ of 0.023 and 0.03 for \gem\ and \psh, respectively.
 These values are consistent with those suggested by previous studies.  
Apparently smaller value of $E(B-V)$ for \gem\ is compatible with the fact
that it is by about 100~pc closer to us than  \psh. 
We applied these excesses
to calculate dereddened fluxes in a standard way making use  
of mean Galactic extinction curve (Cardelli \etal~\cite{card89}) and
the de-reddening parameter $R(V)=3.1$.
The results are presented in Tables~\ref{t:633_fl} and \ref{t:656_fl}.
For both pulsars the dereddening corrections  are  within $1\sigma$ of the flux  
uncertainties in the optical-NIR range, while they 
are significant in the UV.

We note that spectral analysis of recent Chandra and XMM 
X-ray observations  with previously used standard spectral models give  
systematically higher $N_H$ values than those of the ASCA and
ROSAT:   $\approx$2.4$\times 10^{20}$cm$^{-2}$ 
for Geminga (Kargaltsev et al.~\cite{Karg05}) and  $\approx$4.3$\times
10^{20}$cm$^{-2}$ for \psh\  (De Luca et
al.~\cite{deluca}).  The latter one corresponds to 
$E(B-V) \approx 0.09$ which   
is comparable with the  extinction  throughout
the entire Galaxy in this direction (Schlegel
et al.~\cite{Schl98}) and hardly can be  
relevant  for the nearby pulsar. 
$N_H$ for Geminga appears to be  also 
overestimated and we skip  out  
these  implausibly  high  
values\footnote{The   
overestimation of $N_H$ in both cases is likely due to
the poorly calibrated instrument responses  in   the energy range of 
$\la 0.3$~keV  (e.g., Kargaltsev et
al.~\cite{Karg05}) which is  most  crucial for   $N_H$
estimates. This also leads to  lower 
temperatures of  thermal emission components  
and to higher spectral indices of  
nonthermal components derived from spectral fits.}  
from our consideration and use the above ones provided by early
ASCA, EUVE, and ROSAT  data analyses.  We note, that  $N_H$ was also 
fixed at the ROSAT value  in  the recent analysis of the Geminga X-ray  
spectrum based on the most complete data set taken with  the XMM 
(Caraveo et al.~2004).
\subsubsection{Broadband spectra from NIR through UV}
\begin{figure*}[t]
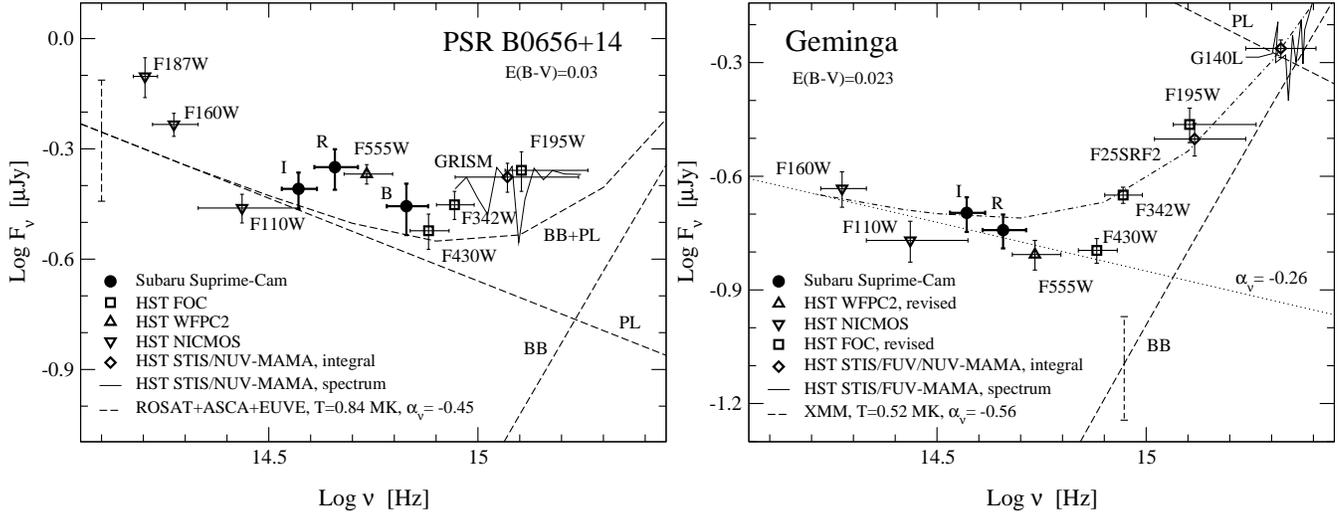

\setlength{\unitlength}{1mm}
\resizebox{12cm}{!}{
\begin{picture}(120,75)(0,0)
\put(0,0){\includegraphics[width=89mm, clip]{fig7a.eps}}
\put(90,0){\includegraphics[width=89mm, clip]{fig7b.eps}}
\end{picture}
}
\caption[h]{Dereddened NIR--UV broadband 
spectra  of  
\psh\  ({\sl left}) and  Geminga ({\sl right})  
summarizing  the best  up to date quality observations obtained with different
telescopes, instruments and filters,  as indicated in the
plots and listed in Tables 5 and 6.  Note  a nonmonotonic 
structure and remarkable similarity  
of the  broadband spectra for  both  pulsars.  
Thin solid curves show  
the spectroscopic data obtained   
 for  \psh\  in the NUV (Shibanov et al.~\cite{Sh2005}) 
 and  for Geminga  in  the FUV (Kargaltsev et al.~\cite{Karg05}).   
Dashed lines  are the low energy extensions of the unabsorbed
soft blackbody (BB) and power law  (PL)
spectral components  derived  from the
combined BB+BB+PL spectral fits 
in X-rays.  Parameters of the fits are taken from Koptsevich et al.~(\cite{K2001}) 
for \psh\ and from
Kargaltsev et al.~(\cite{Karg05}) for Geminga.   
The second, hard  BB  components   are
negligible for both pulsars in the NIR--UV  range and not shown in the plots. 
 Dotted and
dot-dashed lines in the {\sl right panel} show the PL fit
of the Geminga optical data in the range from
$F160W$ through $F430W$ bands and its sum with the BB,
respectively (see Sect.~4.1.3 for details). 
}
\label{f:MAPsp}
\end{figure*}

The dereddened broadband spectral energy distributions of the two pulsars are shown in  
Fig.~\ref{f:MAPsp}. Despite a factor of 2--3  difference  between  the mean
flux levels of the two objects and higher flux   uncertainties in case of Geminga,  
the spectral shapes   of the optical emission of both middle-aged pulsars are  remarkably similar 
to each other.   Both are clearly nonmonotonic with  apparent dips in the $F430W$ and $F110W$ bands  
and  excesses centered in the $V$-$I$ range. \psh\ has also a significant 
flux increase towards  lower frequencies in the NIR range. Similar increase is  
seen in the spectrum of \gem\ but it is less significant due to the absence 
of observations in the $K$  band (or in its HST analog 
$F187W$).  There is also flux  increase  towards  higher
frequencies  in the UV range which is most prominent for Geminga  
where the FUV data are available. 
 The spectroscopic data recently obtained    
 for  \psh\  in the NUV (Shibanov et al.~\cite{Sh2005}) 
 and  for Geminga  in  the FUV (Kargaltsev et al.~\cite{Karg05})
 are also shown in Fig.~7 for comparison. 
 They are compatible with the broadband data, 
 but the flux uncertainties (not shown) in spectral
 bins are  considerably larger than photometric ones, and  apparent 
 absorption/emission features  in the spectra  have a 
 low significance of  $\la2\sigma$. 
 
Resulting spectral energy distributions cannot be fitted
by a single power-law, as it is possible for the young Crab pulsar in the
whole observed range  (Sollerman \cite{S03}; Sollerman et al.~\cite{SLL00}).
This suggests a multicomponent structure of the emission of 
the two middle-aged pulsars and may reveal  spectral evolution of the optical emission 
with pulsar age (K\cite{K2001}; Shibanov et al.~\cite{Sh2003};  Zharikov et
al.~\cite{Z2004}). 
Some of the components can be identified from the comparison with the X-ray data.  
\subsubsection{Thermal and nonthermal spectral components} 
\label{s:th-nth}
For  PSR B0656+14   the flux increase with frequency  in  the UV    
can  be   explained by a contribution of the  Rayleigh-Jeans part of  
the thermal blackbody-like spectral component (BB) from the whole surface of
the NS strongly dominating  the pulsar emission 
in soft X-rays. The flux increase  towards the infrared  
appears to be consistent with the low energy extension of
the power law  component (PL) seen in the high energy tail
 of  the X-ray spectrum. The unabsorbed PL and BB components and their sum are  
 shown by dashed lines in Fig.~7. Their parameters 
are taken from K\cite{K2001}: the BB temperature $T_{BB}$$\approx$0.84~MK
and the radius of the emitting area $R_{BB}$$\approx$7.8~km at
the distance $d$$\approx$288~pc; the PL spectral index $\alpha_{\nu}$$\approx$-0.45 and normalizing constant
at 1~keV is  $\approx$2.8$\times$10$^{-5}$~ph cm$^{-2}$ s$^{-1}$
keV$^{-1}$;  $N_H$$\approx$1.42$\times 10^{20}$~cm$^{-2}$. 
The $N_H$ value and  $E(B-V)$ in the plot are consistent with each other  
as discussed above.  Uncertainties of the X-ray spectral 
fit extension are shown by a dashed errorbar.    
As seen from the {\sl left panel} of Fig.~7, 
the BB+PL extension   qualitatively fits the    
spectrum of \psh,  albeit the PL component dominates in the most part 
of the observed optical range.  This is also confirmed by detailed 
analyses of the pulsar pulse profile and the spectrum  in the NUV, where
the BB contribution was estimated  to be $\la30\%$ 
(Shibanov et al.~\cite{Sh2005}).     

For Geminga the situation is more complex.  
The parameters of the X-ray extensions  in this case  
 are taken from Kargaltsev et al.~(\cite{Karg05}): 
$T_{BB}$$\approx$0.52~MK, $R_{BB}$$\approx$9.5~km at 
$d$$\approx$200~pc; the PL index $\alpha_{\nu}$$\approx$-0.56 and  
normalizing constant at 1~keV is  $\approx$5.5$\times$10$^{-5}$~ph cm$^{-2}$ s$^{-1}$  
keV$^{-1}$;  $N_H$ was fixed during the X-ray spectral fit 
at the value of 1.1$\times $10$^{20}$~cm$^{-2}$. 
 As seen from the {\sl right panel} of Fig.~7,  the  PL and, therefore, 
 the sum BB+PL (it is outside the upper plot frame)        
overshoot  the optical fluxes of Geminga by a factor of $\ga100$.  
This fact has been noticed  by Halpern and Wang~(\cite{hal97}) and
confirmed later on by Jackson et al.~(\cite{jak02}), Komarova et al.~(\cite{K2003}), and Kargaltsev et
al.~(\cite{Karg05}).  At the same time,  the BB extension appears 
to fit qualitatively well a steep flux increase of Geminga towards higher  
frequence  range. 

Various possibilities to resolve the discrepance between 
the low  energy extension of the Geminga X-ray spectrum 
and the observed optical fluxes       
have been  discussed  by Kargaltsev et al.~(\cite{Karg05}), 
including possible systematic errors due to instrument response 
uncertainties, large statistical uncertainties of the X-ray  
PL component, its contamination by 
the hard BB component from the pulsar  polar caps,  and possible existence 
of  a spectral break in the spectrum of the nonthermal  component
somewhere between the optical and X-ray ranges. 

The similarity of
the  optical spectral shape of Geminga with that of  PSR B0656+14 
suggests that its low frequence tail is of nonthermal
origin and can be fitted by a power law. Such a fit
performed for  the $F160W$--$F430W$  range  
is shown by a dotted line in the {\sl right panel} of Fig.~7.  
Its sum with the  BB extension (dot-dashed line)   
qualitatively fits  the whole set of the data in the NIR--UV range, 
as it is in case  of the BB+PL extension for  PSR B0656+14.     
The slope of the optical PL,  $\alpha_{\nu}$$\approx$-0.26, is almost  
twice smaller than that  in X-rays,  $\alpha_{\nu}$$\approx$-0.56, 
favoring  the interpretation with the spectral  break in the nonthermal  
optical-X-ray emission of Geminga.  
Changing  the spectral slope of the nonthermal component 
from almost a flat in the optical to a significantly negative in X-rays is  not unusual 
and has been observed in the emission of the young  Crab pulsar 
(Kuiper et al.~\cite{Kuip01}; Sollerman~\cite{S03}), older Vela (Shibanov et al.~\cite{Sh2003}; 
Romani et al.~\cite{romani05}), and likely in two old pulsars 
PSR B1929+10 and PSR B0950+08 detected in the optical 
(Pavlov et al.~\cite{PSC}; Mignani et al.~\cite{mig02};  Zharikov et al.~\cite{Z2002}; \cite{Z2004})  
and X-rays (Becker et al.~\cite{becker04} and \cite{becker05};
Zavlin \&  Pavlov~\cite{zp04b}). The nonthermal emission of very old isolated millisecond pulsars 
studied in the optical and X-ray ranges shows even a much stronger 
slope break between these ranges (Koptsevich et al.~\cite{Kopts03}; 
Mignani \& Becker~\cite{mig04}). We discuss the possible position of the break 
for Geminga below in  Sect.~4.2.  
\subsubsection{Third optical spectral component?} 
\label{s:3d-comp}
A wide $F555W$-$R$ excess  in the emission of PSR B0656+14 
cannot be explained by only the combination of  the  Rayleigh-Jeans and  power-law components.
It exceeds by  $\sim (3-5)\sigma$  a power-law ``continuum level'' supposed to be seen in the $F430W$ and $I$ bands. 
The excess may reveal a new, not yet identified,  {\it third  spectral component} in the emission of \psh.  

The nature of the excess is unclear.  
One can speculate that it is produced by unresolved {\it  emission lines } 
from a compact  PWN around the pulsar interacting 
with  interstellar environment polluted by heavy 
elements from the SN explosion associated with
the pulsar. Strong OIII (5007\AA), SII (6717\AA) and other fainter nebular lines, whose  
wavelengths  are  within the spectral range of the excess, 
are  typical for the spectra  of bright and extended 
PWNe formed around  young  pulsars like Crab. 
In case of PSR B0540-69 a strong background 
emission in OIII (5007\AA) line from the SNR core contaminates the  pulsar photometric 
flux in the $F555W$ band and results in a false  flux excess  exceeding  
by an order of a magnitude the real  flux from the pulsar 
(Serafimovich et al.~\cite{ser04}). 
A distant  outer shell of the \psh\ SNR    
has been detceted  in X-rays     (Nousek et al.~\cite{nousek81}; Thorsett et al.  \cite{thor03}).   
 However,  nothing is known on the presence of the nebular 
lines in the pulsar vicinity,  although  they could    exist 
if the pulsar PWN marginally detected in the optical  (Fig.~6) and X-rays 
(Marshall \& Schulz \cite{Mar02}) is real.

Alternatively, the excess  can be 
an internal property of the pulsar emission. 
It can be produced by {\it ion or electron cyclotron features} formed 
in magnetospheric plasma  of a strongly magnetized NS. 
The features are expected to  be broad enough due 
to the  magnetic field inhomogeneity at the magnetospheric 
spatial scale. Current data do not allow us to discriminate between   
the two  possibilities. 
To do that a deep {\it narrow-band imaging } of the pulsar field
with   a standard set of narrow-band filters centered 
at  nebular lines and/or   {\it spectroscopy} of the pulsar  are necessary. 
          
\gem\  is fainter and  its fluxes are determined with larger 
errors.  The apparent I-R  
excess  in its spectrum is much less  significant than for PSR B0656+14.  However, the similarity of the 
spectral shapes of  both pulsars provides an additional argument that this excess is real.     
This is also supported  by the presence of the X-ray  nebulosity around \gem\ and its possible 
optical counterpart (Fig.~5), which may be resposible for
the excess. In this case, one can suggest the same two 
possibilities to explain its nature as has been discussed above for \psh.

We note, that with the new  optical and NIR data the  optical spectrum of  Geminga 
cannot be described by a two component model  (Jacchia et al. \cite{Jacchia})  
combining only an ion cyclotron emission line  
from a hot   plasma near the polar caps of  the pulsar and the thermal 
radiation from  the cooling surface of the NS.  
The nonthermal PL spectral component has to be included into 
the interpretation and it strongly dominates  the emission
at longer wavelengths.  
\begin{figure*}[t] 
\centering
\includegraphics[width=13.8cm, bb=132 145  510
716]{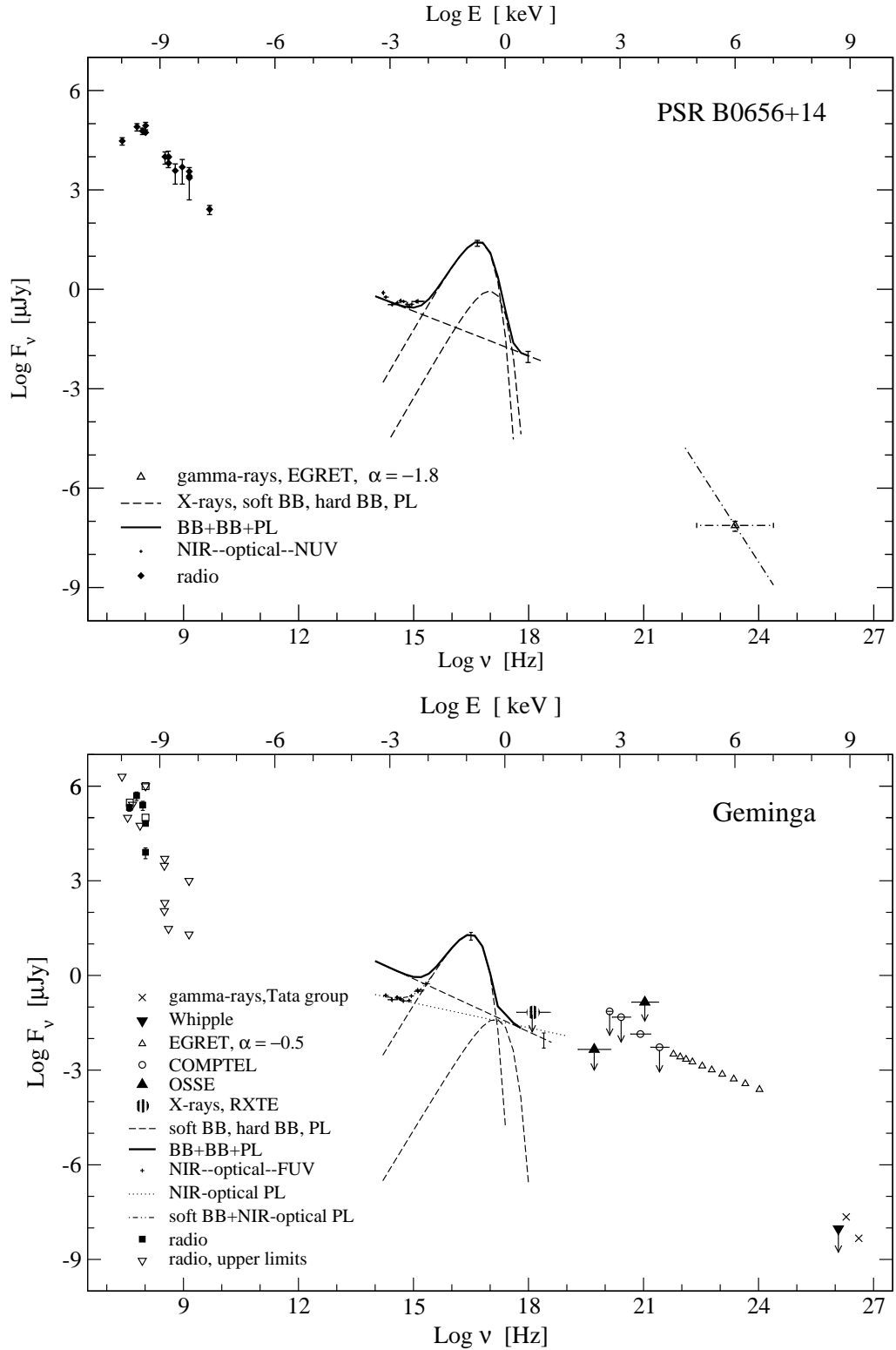}
\caption{
Unabsorbed multiwavelength spectra of the two
pulsars  
using available data  
from radio through gamma-rays, as notified in the plot. X-ray spectral
components are extended to the optical range, while the optical one -- to X-rays.  
}
\label{f:633-656-multi}
\end{figure*}  
\subsection{Multiwavelength spectra}
\label{s:mw}
In Figure~\ref{f:633-656-multi}  we show unabsorbed    rotation-phase integrated  multiwavelength 
spectra of both pulsars using  available data from the radio through  gamma-rays.  The radio data for Geminga are kindly 
provided by V.~Malofeev (private communication). There are some doubts  
on the  radio detection  (Burderi et al~1999)  and the  data are   represented
mainly by upper limits.    
For PSR B0656+14 the radio data are  from K\cite{K2001}. 
The optical data are from Tables 5 and 6.   
The parameters of the X-ray spectral 
components are the same as in Fig.~7. For both NSs we add here 
also the hard BB components believed to be from 
the pulsar polar caps. 
They contribute significantly to the resulting spectra  
only near $\sim$1~keV where the soft BB components 
meet the PL tails. Parameters of the hard BBs  are  from K\cite{K2001} 
for  PSR B0656+14 ($R$$\approx$0.52~km at
$d$$\approx$288~pc, $T$$\approx$1.65~MK) 
and from Kargaltsev et al.~(\cite{Karg05}) for Geminga
($R$$\approx$0.046~km at 
$d$$\approx$200~pc, $T$$\approx$2.32~MK).  All X-ray components are extended 
to the optical. The gamma-ray data are from Ramanamurthy et al.~(\cite{r96})  
for \psh\ and  from Jackson et al.~(\cite{jak02}; RXTE), Macomb and Gehrels 
(\cite{mg99}; OSSE, COPMTEL, EGRET)\footnote{\href{http://cossc.gsfc.nasa.gov/docs/cgro/gamcat/} 
cossc.gsfc.nasa.gov/docs/cgro/gamcat/},  Akerlof et al.~(\cite{A93}; Whipple) and  
Vishwanath et al.~(\cite{V93}; Tata group) for Geminga.

The multiwavelength spectra underline  the fact that 
a smooth connection of the optical and X-ray nonthermal and thermal 
spectral components observed for \psh\  is problematic for \gem. 
 \gem\  is obviously under-luminous  in the optical range 
as compared with  \psh\ and with the low-energy extension of 
its power-law component from  X-rays. 
This suggests  changing the spectral slope  of 
the nonthermal component of \gem\ from a significantly negative in X-rays 
($\alpha_{\nu}$$\approx$-0.56, dashed line) 
to almost a flat ($\alpha_{\nu}$$\approx$-0.26, dotted line) 
in the NIR-optical range. A simple extrapolation of the NIR-optical PL fit 
to X-rays shows  that the spectral break in the nonthermal optical-X-ray 
emission may occur near 1 keV where the hard BB component dominates the total
flux ({\sl bottom panel} of Fig.~8). 

A common origin of the optical and X-ray nonthermal 
components  independently  on  pulsar age and the presence of spectral breaks   
was proposed by Zharikov et al.~(\cite{Z2004}; \cite{Z2005}) 
based on  a strong correlation of the optical and X-ray 
nonthermal luminosities for a set of ordinary pulsars detected in the optical 
and X-ray domains (see also Zavlin \&  Pavlov~\cite{zp04b}). 
Analysis of the phase aligned optical 
and X-ray pulse profiles  would be useful to verify this hypothesis 
for Geminga, as it has been partly done for \psh\ 
(Shibanov et al.~\cite{Sh2005}). Detection of the optical pulsations of Geminga 
at a higher $S/N$  than it is provided by a tentative detection 
by Shearer et al.~(\cite{Shear98})  is necessary for this goal.

The multiwavelength nonthermal 
spectrum  of  \gem\  suggests also at least three additional  
spectral breaks in the high energy range: in hard X-rays,  
in  gamma-rays, and in the TeV range. 
They   form a  ``knee-like''  spectral structure  of 
the high energy emission. Similar ``knee''  is seen in the high energy 
spectrum of the Vela-pulsar (e.g., Shibanov et al.~\cite{Sh2003};  
Romani et al.~\cite{romani05} and references therein). The optical part of 
the Vela spectrum suggests also a ``degenerate knee''  
between the optical and X-rays  while its NIR and X-ray data 
from the nonthermal spectral tail can be fitted by a single power-law 
(Shibanov et al.~\cite{Sh2003}). It is not clear whether the optical-X-ray and  
X-ray-gamma-ray  knees are correlated or physically related  to  each other.  
 The Crab pulsar does not have such a strong spectral breaks 
  and knees in the high energy range, while it has a strong spectral 
  break between the optical and X-rays (Kuiper et al.~\cite{Kuip01}; Sollerman \cite{S03}). 
  The same appears to be true for PSR B0540-69 (Serafimovich et al.~\cite{ser04}).          
Various slopes and numerous breaks likely suggest a multicomponent 
structure of the emission generated by different mechanisms or 
by different populations of the relativistic particle, or both. 
 However,  multiwavelength data for  pulsars 
 are too scarce to make any definite conclusion. 
 For instance, PSR B0656+14 appears to have no significant 
 spectral break between the optical and X-rays, 
 while almost nothing is known on its high energy spectrum  
 except  it is likely  softer than that of Geminga.

The problem of the spectral difference of the pulsars outside 
the radio range is far from its final solution. 
A question, whether  this difference simply caused  by different 
pulsar emission beaming and geometry factors or by  significantly different  
physical  conditions in the pulsar magnetospheres, is not answered yet. 
Studying  the pulse profiles in different  spectral ranges 
and phase resolved spectroscopy can help  to solve this problem and,
particularly, to understand  whether the optical and X-ray nonthermal photons 
are generated  at the same site and by the same population of 
relativistic particles  accelerated in the pulsar magnetospheres. 
If the latter is true,  then one has to understand  
why the distribution function of the particles in some cases reveals 
a break at an intermediate energy, which is reflected in 
the observed  spectra. 
\subsection{Possible optical PWNe}   
 The sizes of axisymmetrical equatorial torus-like (plus polar jets) 
 PWNe  around  the young  Crab, PSR B0540-69, and Vela  
 pulsars are roughly  scaled  as $\dot{E}^{0.5}$, where  
 $\dot{E}$ is the spin-down luminosity of a pulsar  
 (Helfand et al.~\cite{hel01}; Pavlov et
 al.~\cite{pav01b}; Serafimovich et al.~\cite{ser04}; Weisskopf et al.~\cite{w04}).  This is in agreement with the simplest  
 model of a termination wind shock forming  around subsonic-ally moving energetic 
 pulsars   due to the interaction of their relativistic  winds with 
  the ambient matter (Kennel \& Coroniti~\cite{kc84}).  
  The PWNe of the Crab and its twin PSR B0540-69 are  extended   
  approximately to about 1~pc from the pulsars and both  have similar sizes 
  in the optical and X-rays (e.g., Serafimovich et al.~\cite{ser04}). Using this value and  the above scaling law  
  one can  estimate that  PWNe around  the middle-aged \gem\ and \psh\  have to be  about a factor 
  of 100 smaller than in  the Crab case, i.e, a few arcseconds  at the distances 
  to the pulsars of about 200-300~pc, provided the external density and 
  pressure conditions are not much different from that of Crab 
  and are appropriate to form the subsonic torus but not a supersonic 
  bow shock. The sizes  of the faint extended  structures 
  around both middle-aged pulsars marginally detected 
  in our deepest  Subaru images are in a reasonable agreement with these 
  expectations. This is a starting point to discuss these structures  as possible 
  candidates for PWNe. 
  
  So far  PWNe of a different nature and shapes have been detected in the radio, 
  optical and X-rays around a dozen of isolated NSs (e.g., Gaensler~\cite{gaen05}). 
   Young NSs typically show  the torus/jet-like PWNe, while the PWNe around old NSs commonly have  
  a bow shock structure.  Changing of the shock type with age is 
  supposed to be due to a  transformation of external conditions from the subsonic 
  to supersonic ones when the density and the pressure within the expanding  host  
  SNR decrease  to  values of a typical ISM.  A characteristic age of a   
  host SNR when it dissolves in the ISM background roughly coincides 
  with  a middle-age epoch  of the NS evolution, i.e., $10^5$ -- $10^6$ yr. 
 
  Following this scenario one can expect to see 
  an intermediate PWN morphology, i.e., a mixture 
  of torus and bow-shock,   around middle-aged pulsars. 
  However, until now there are only  two middle-aged pulsars  
  with a faint, few arcseconds  PWNe apparently  detected  only in the radio 
  around PSR  B0906-49 (Gaensler et al.~\cite{gaen98}) and only in X-rays 
   around PSR J0538+2817 (Romani \& Ng~\cite{romani03}). Both cases need additional 
   observations in different wave bands to confirm  the nature 
   of the extended objects around the pulsars and to study  
   their morphology.  In this respect, adding to this poor sample any detection of PWNe, 
  even a tentative,  would be valuable.

  In contrast to a  bright  Crab PWN, whose contribution to the total 
  pulsar+nebula luminosity is about 95.5\% and 99.8\%  in X-rays and 
  in the optical,   respectively, the PWN-like structures around our middle-aged pulsars 
  are faint and may contribute only a few percents to the  total   emission budget. 
 However, this is compatible with  the observational fact that the torus PWNe 
 loose  their  brightness  with age much sooner than the pulsars. 
 For instance, the PWN contribution for an intermediate  age  
 Vela-pulsar ($10^4$~yr)   in X-rays is  about  75\% (Pavlov et al.~\cite{pav01b}). It is likely to be much 
 smaller in the optical range where  the Vela-PWN has not yet been clearly identified (Shibanov et al.~\cite{Sh2003};
 Mignani et al.~\cite{mig03}). A faint X-ray PWN  tentatively detected around 
 a middle-aged pulsar PSR J0538+2817  ($6\times 10^5$~yr) provides  only about 2\% of the pulsar flux 
 in X-rays (Romani \& Ng~\cite{romani03}).   This is in agreement  with  what is seen  in  our case of \gem\ and \psh.              
                   
  Another factors suggesting that the extended optical
  structures  around our pulsars 
  are likely to be  PWNe are their morphology and orientation to 
  the vectors of the pulsar proper motion.      
  The planes of the equatorial torus structures of  young pulsars 
  are typically arbitrary inclined to the line of sight and
  seen as elliptical extended objects.  
   For the Crab, Vela, and, possibly, PSR B0540-69 pulsars  
   the torus symmetry (pulsar spin)  axis  and PWN polar jets   
   are  aligned with the vector of proper motion of the pulsar, 
   which  may constrain  the nature of the kick  
   at pulsar birth (Serafimovich et al.~\cite{ser04}).  Both properties are  
   likely seen for \gem\ and \psh\ where the extended optical structures 
   (possible inclined tori) are aligned perpendicular 
   to the pulsar proper motion, although  any jet/counter-jet structures  
   are not visible  in our images.  
     
    Finally, possible pulsar nebulae around  \gem\ and \psh\ have been detected   in  X-rays,   
   albeit at different significance levels.           
   The sizes  of a less significant circle  X-ray nebulosity around \psh\ 
   (Marshall \& Schulz~\cite{Mar02})   
   are consistent with what is seen in the Subaru image.  
     \gem\ shows apparently more significant  long 
     (up to $\sim$2\arcmin) 
    X-ray tails well aligned with the source proper motion 
    (Caraveo et al.~\cite{Car03}). 
    The tails evidence that the PWN of the older \gem\   
    has   transformed  from the subsonic torus to the supersonic  
    bow shock stage. This is also supported by a twice higher transverse 
    velocity of \gem, by absence of its host SNR, and, hence, 
    likely smaller density and external pressure of the ambient matter   
    than in case of \psh. The latter one is younger and  still sits inside 
    its host SNR (Thorsett et al.~\cite{thor03}). 
    Note also, that its DM is significantly  and $N_H$  
    is apparently  higher than in the \gem\  case, favoring  a denser  
    environment for \psh.  Consequently,  an  arc-like   shape of 
    the faint extended structure at the \gem\  optical image suggests 
    that it is likely to be  a counterpart of the head of the bow-shock 
    (but not a bright part of a PWN torus) whose tails are detected in X-rays.   
    
    Thus, the  morphology of both  optical  nebulosities tentatively detected 
     around the two middle-aged pulsars  are roughly  compatible 
     with what is expected  
    from their parameters and parameters of the environments, as well as with 
    what is seen for each pulsar in X-rays. Confirmation of their reality  
    at higher significance level and with better spatial resolution  
    in the optical and X-rays would be valuable for understanding   
    the  physics of PWNe and their evolution with age.    
\acknowledgements 
We are grateful to Yutaka Komiyama 
for the help during observations with the Subaru, 
to Valeri~Malofeev for  
radio data for Geminga, and to Roberto Mignani for the ESO/NTT optical 
data and stimulating comments. This work was supported in part by CONACYT 
project 25454-E, RFBR (grants 03-02-17423, 03-07-90200 and 05-02-16245) 
 and RLSS programma 1115.2003.2.  
Some of the data presented here were obtained from the
Multi mission Archive at the Space Telescope Science Institute (MAST). 
STScI is operated by the Association of Universities for Research in 
Astronomy, Inc., under NASA contract NAS5-26555. 
Support for MAST for non-HST data is provided by the NASA Office 
of Space Science via grant NAG5-7584 and by other grants and contracts.


\begin{thebibliography}{}

\bibitem[1993]{A93}
Akerlof, C. W., Breslin, A. C., Cawley, M. F., et al. 1993, \aap, 274, L17

\bibitem[1997]{Becker}Becker W., \&  Tr\"{u}mper J. 1997
\aap , 326, 682

\bibitem[2004]{becker04} Becker W.,    Weisskopf, M. C.,  Tennant, A. F.,
et al. 
2004, \apj, 615, 908


\bibitem[2005]{becker05} Becker W., Jessner, A., Kramer,
M.,   
Testa, V., Howaldt, C. 
2005, \apj (accepted, astro-ph/0505488) 
                     
\bibitem[1996]{big96}  Bignami,  G.F. \&  Caraveo, P.A. 1996,
\araa, 34, 331 

\bibitem[1996]{bcmeb96}
Bignami, G.F., Caraveo, P.A., Mignani, R.,  
 Edelstein, J.,  Bowyer, S. 1996,
\apj,  456, L111



\bibitem[1978]{boh1978}
Bohlin R.C., Savage B.D.  Drake, J.F. 1978, \apj, { 224}, 132

\bibitem[2003]{brisk03} Brisken, W F., Thorsett, S. E.,
Golden, A., Goss, W. M. 2003, \apj,{ 593}, L89

 \bibitem[1999]{Bur99} Burderi,~L., 
Fauci,~F., Boriakoff,~V.  1999, ApJ, 512,  L59  

\bibitem[1996]{Car96}
Caraveo, P. A., Bignami, G. F.,  Mignani, R.,   
 Taff, L. G.
 1996, \apj, 461, L91

\bibitem[1998]{Car98}Caraveo, P. A., Lattanzi, M. G., Massone, G., et al. 1998, \aap, 329, L1

\bibitem[2003]{Car03}
Caraveo, P.A., Bignami, G.F.,  De Luca, A., et al.  2003, Science, {301},
134

 \bibitem[2004]{Car04}
Caraveo, P.A.,  De Luca, A., Mereghetti, S.,   
 \& Bignami, G.F. 
2004, Science, 305,
376

\bibitem[1989]{card89}
Cardelli, J. A., Clayton, G. C., Mathis, J. S. 1989, \apj,  345, 245


\bibitem[1993]{cord93} Cordes, J. M., 
    Romani, R. W., 
    Lundgren, S. C. 1993, \nat, 362, 133 


\bibitem[2005]{deluca}De Luca, A., Caraveo, P. A.,
Mereghetti, S.,  
 Negroni, M., Bignami, G. F. 
 2005, \apj, 623, 1051



\bibitem[1998]{Fruchter} Fruchter, A.S., Hook, R.N.,
Busko, I.S.,   
  Mutchler, M. 
  1998, in 1997 HST Calibration Workshop, eds.
S. Csertano, R. Jedrzejewsky, T. Keyes, and M. Stevens (STScI)

\bibitem[1995]{Fukugita}
Fukugita, M., Shimasaku, K., Ichikawa, T. 1995,
{\pasp,} { 107}, 945



\bibitem[1997]{hal97} Halpern J.P., \& Wang, Y.-H. 1997, \apj, { 477}, 905



\bibitem[1998]{Harlow}
Harlow J.J.B., Pavlov G.G., Halpern J.R. 1998,
{\it AAS Meeting\/} 193, \#41.07


\bibitem[2001]{hel01} Helfand, D. J., Gotthelf, E. V., Halpern, J. P.
2001, \apj, 556, 380L 


\bibitem[2002]{hes02} Hester, J. J., Mori, K., Burrows, D., et al.
2002, \apj, 577, L49 

\bibitem[1998]{gaen98} Gaensler, B. M., Stappers, B. W.,
    Frail, D. A.,  
     Johnston, S. 
    1998, \apj, 499, L69
    

\bibitem[2005]{gaen05} Gaensler, B. M. 2005, ASpR, 35, 1116  



\bibitem[1999]{Jacchia}Jacchia, A., de Luca, F., Lazzaro, E.,
et al.
1999, \aap, 347, 494


\bibitem[2002]{jak02} Jackson, M. S., Halpern, J. P.,
Gotthelf, E. V.,  
 Mattox, J. 
2002, \apj, 578, 935 

\bibitem[2005]{Karg05} Kargaltsev, O. Y.,
 Pavlov, G. G., Zavlin, V. E., Romani, R. W. 2005, \apj, 625, 307  

\bibitem[1999]{Kassim} Kassim, N., \& Lazio, T.J. 1999, \apj, 527, L101


\bibitem[1984]{kc84} Kennel, C. F., \& Coroniti, F. V.
   1984, \apj, 283, 694


\bibitem[2003]{Kern2003}
Kern, B., Martin, C., Mazin, B., Halpern, J. P.
\apj, 597, 1049

\bibitem[2003]{K2003}
Komarova, V., Shibanov, Yu., Zharikov, S., 
\etal 2003, 
Pulsars, AXPs and SGRs observed with BeppoSAX and Other Observatories,
Proceedings of the International Workshop held in Marsala, September 23-25,
2002. Edited by G. Cusumano, E. Massaro, T. Mineo,  77



\bibitem[2005]{K2005}Komarova et al.~2005 (in preparation) 


\bibitem[2001]{K2001}Koptsevich A., B., Pavlov G. G.,
Zharikov S. V.,
et al.
2001 (K2001), \aap, 370, 1004 

\bibitem[2003]{Kopts03} Koptsevich, A. B.,
   Lundqvist, P., 
   Serafimovich, N. I.,   
 Shibanov, Yu. A., 
  Sollerman, J. 
 2003, A\&A, 400, 265



\bibitem[2001]{Kuip01} Kuiper, L., Hermsen, G., Cusumano, G., 
   et al. 2001, \aap, 378, 918


\bibitem[1998]{Kurt98}
Kurt, V.G., Sokolov, V.V., Zharikov, S.V.,  
Pavlov, G.G.,  Komberg, B.V. 
1998,
\aap, { 333}, 547

\bibitem[2001]{Kurt2001} Kurt, V.G., Komarova, V.N., Sokolov, V.V., \etal
2001, BSAO, {51}, 21

\bibitem[1997]{kuz97} Kuz'min, A.D., \& Losovskii, B.Ya. 1997,
Astronomy Letters, 23, 283

\bibitem[1992]{lan92}
Landolt A.U. 1992, \aj, { 104}, 340


\bibitem[1999]{mg99} 
Macomb, D.J., \& Gehrels, N. 1999, \apjs, 120, 335 

\bibitem[1997]{MM97}
Malofeev V.M., \& Malov O.I. 1997, \nat, {389}, 697

\bibitem[2002]{Mar02} Marshall, H. L., \& Schulz, N.S. 2002, \apj,
{574}, 377

\bibitem[1998]{Mar98}
Martin C., Halpern J.P., Schiminovich D. 1998, \apj, {494}, L211


\bibitem[1998]
{mcb98} Mignani, R.P., Caraveo P.A.  Bignami G.F. 1998 \apj, 332, L37

\bibitem[2000] {Mig2000}
Mignani, R.P., De Luca, A.,  Caraveo, P.A. 2000,
\apj, {543}, 318

\bibitem[2002]{mig02} Mignani, R. P., 
     De Luca, A., 
     Caraveo, P. A.,   
     Becker, W. 
     2002, \apj, 580, L147  

\bibitem[2003]{mig03} Mignani, R. P., De Luca, A., 
     Kargaltsev, O., et al. 2003, \apj, 594, 419 

\bibitem[2004]{mig04} Mignani, R. P., 
\&     Becker, W. 2004, AdSpR, 33, 616 



\bibitem[1998]{Miyazaki}
Miyazaki, S., Sekiguchi, M., Imi, K.,
et al.
1998,
in Proc. SPIE 3355: Optical Astronomical
Instrumentation, ed. S. D'Odorico, 363

\bibitem[2002]{Monet}
Monet, D., Levine, S., Canzian, B., et al.  
 2002, \baas, 34, 1104 (astro-ph/0210694)

\bibitem[1981]{nousek81} Nousek, J. A., Cowie, L. L., Hu, E.,   
Lindblad, C. J., Garmire, G. P. 
1981, \apj, 248, 152 

\bibitem[1996]{PSC} 
Pavlov, G.G., Stringfellow, G.S., C\'ordova, F.A. 1996,
\apj, { 467}, 370

\bibitem[1997]{PWC}
Pavlov, G.G., Welty, A.D., C\'ordova, F.A. 1997,
\apj, {489}, L75

 \bibitem[2001]{pav01b} Pavlov, G. G., Zavlin, V. E.,
   Sandwal, D. et al. 2001, \apj, 554, 189


\bibitem [2002]{Pav02} Pavlov, G.G., Zavlin, V.E., Sanwal D. 2002,  
Proceedings of the 270. WE-Heraeus Seminar on Neutron Stars, 
Pulsars, and Supernova Remnants, 273  (astro-ph/0206024)

\bibitem[2005]{pell05}Pellizza, L. J., Mignani, R. P., 
Grenier, I. A.,   Mirabel, I. F.   
2005, \aap, 435,
625 

\bibitem[1996]{r96}
Ramanamurthy, P.V., Fichtel, C.E., Harding, A.K., et al.
 1996,
\aaps, 120, 115
 \bibitem[1997]{romani97} Romani, R. W., 
   Cordes, J. M., 
   Yadigaroglu, I.-A.~1997, \apj, 484, L137  

 \bibitem[2003]{romani03}  
Romani, R.~W. \& Ng, C.-Y. 2003, \apj, 585, L141

 \bibitem[2005]{romani05} 
Romani, R.~W.,  Kargaltsev, O., Pavlov, G.~G.  
2005, \apj, 627, 383


\bibitem[1998]{Schl98}
Schlegel, D.J., Finkbeiner, D.P., Davis, M. 1998, \apj,  {500}, 525

\bibitem[2004]{ser04} Serafimovich, N.~I., Shibanov, Yu.~A.,
Lundqvist, P.,   
Sollerman, J.
~2004, \aap, 425, 1041 


\bibitem[1997]{Shear97}
Shearer, A., Redfern, R.M., Gorman, G., \etal
1997, \apj, 487, L181

\bibitem[1998]{Shear98} Shearer, A., Golden, A., Harfst, S., \etal
1998, \aap, {335}, L21



\bibitem[2003]{Sh2003}
Shibanov, Yu. A., Koptsevich, A. B., Sollerman, J.,  Lundqvist, P. 
 2003,
\aap, 406, 645

\bibitem[2005]{Sh2005} Shibanov, Yu. A.,
 Sollerman, J., Lundqvist, P., 
Gull, T., Lindler, D.
 2005, \aap, 440, 693  

\bibitem[1998]{Shi98} Shitov Yu.P., \& Pugachev V.D. 1998, New Astronomy,
{3}, 101

\bibitem[2000]{SLL00} Sollerman, J., Lundqvist, P., 
Lindler, D., et al. 2000, \apj, 537, 861

\bibitem[2003]{S03} Sollerman, J. 2003, \aap, 406, 639

\bibitem[1993]{Taylor}
Taylor, J.H., Manchester, R.N.,  Lyne, A.G. 1993,
\apjs, 88, 529.

\bibitem[2003]{thor03}
Thorsett, S. E., Benjamin, R. A., Brisken, W., \etal
2003, \apj, {592}, L71


\bibitem[1993]{V93}
Vishwanath, P.R., Sathyanarayana, G.P., Ramanamurthy, P.V., 
Bhat, P.N. 
1993, \aap, 267, L5


\bibitem[2004]{w04} Weisskopf, M. C, O'Dell, S. L.,
   Paerels, F., et al. ~2004, \apj, 601, 1050



\bibitem[2004]{yp04} Yakovlev, D. G., \& Pethick, C. J. 2004, 
 \araa, 42, 169 

\bibitem[2004a]{zp04a} Zavlin, V. E., Pavlov, G. G. 2004a, 
Memorie della Societa Astronomica Italiana, 75, 458   

\bibitem[2004b]{zp04b} Zavlin, V. E., Pavlov, G. G. 2004b, \apj, 
616, 452 


\bibitem[2002]{Z2002}
Zharikov, S. V., Shibanov, Yu. A., Koptsevich, A. B., et al.
 2002, \aap, 394, 633

\bibitem[2003]{Z2003}
Zharikov, S., Mennickent, R., Shibanov, Yu., 
\etal~2003, 
Pulsars, AXPs and SGRs observed with BeppoSAX and Other Observatories,
Proceedings of the International Workshop held in Marsala, September 23-25,
2002. 

\bibitem[2004]{Z2004}
Zharikov, S. V., Shibanov, Yu. A., Mennickent, R. E. 
\etal 2004, \aap, 417, 1017

\bibitem[2005]{Z2005}
Zharikov, S. V., Shibanov, Yu. A., Komarova, V., N. 
2005, AdSpR,  in press (astro-ph/0410152) 

\end{thebibliography}
\end{document}